\documentclass[aps, twocolumn, amsmath, amssymb, prb, 10pt, floatfix]{revtex4-2}

\newif\ifincludefigures
\includefigurestrue 
\ifincludefigures\fi

\usepackage[per-mode=symbol,separate-uncertainty]{siunitx}

\usepackage[]{graphics} 
\usepackage{amsmath,color,colordvi}  
\usepackage{mathtools}
\usepackage[pdftex]{graphicx} 
\usepackage[english]{babel}   
\usepackage{braket}

\definecolor{navyblue}{rgb}{0.0, 0.0, 0.5}
\usepackage[protrusion=true, expansion=true]{microtype} 
\usepackage[colorlinks=true, breaklinks=false, linkcolor=navyblue, urlcolor=navyblue, citecolor=navyblue]{hyperref}



\makeatletter
\renewcommand{\figurename}{\textbf{Figure}}
\renewcommand{\tablename}{\textbf{Table}}

\renewcommand{\fnum@figure}{\textbf{\figurename~\thefigure}}
\renewcommand{\fnum@table}{\textbf{\tablename~\thetable}}
\makeatother

\begin{document}

\title{Autonomous quantum heat engine}
% \title{Autonomous quantum heat engine based on optomechanical-like coupling with superconducting circuits}

\author{Tuomas Uusn\"{a}kki$^1$}
\email{tuomas.uusnakki@aalto.fi}
\author{Miika Rasola$^{1,4}$}
\author{Vasilii Vadimov$^1$}
\author{Priyank Singh$^{1,3}$}
\author{Ahmad Darwish$^1$}
\author{Mikko M\"{o}tt\"{o}nen$^{1,2}$}
\email{mikko.mottonen@aalto.fi}

\affiliation{$^1$QCD Labs, QTF Center of Excellence, Department of Applied Physics, Aalto University, P.O. Box 13500, FI-00076 Aalto, Finland\\
  $^2$VTT Technical Research Centre of Finland, QTF Center of Excellence, P.O. Box 1000, FI-02044 VTT, Finland\\
 $^3$\rm Present address: \it Center for Quantum Science and Technologies, Indian Institute of Technology, Mandi, Himachal Pradesh, 175075, India\\
 $^4$\rm Present address: \it Arctic Instruments, Innopoli 2, Tekniikantie 14, 02150 Espoo, Finland}
 
\date{March 16, 2026}

\begin{abstract}
Quantum heat engines provide attractive means in quantum thermodynamics for studying the fundamentals of the flow of heat and work. Previous experimental implementations of heat engines operating at the level of a few excitation quanta have utilized external driving, which has made the observation of the produced work challenging. Conversely, autonomous quantum heat engines only require a flow of heat to operate and generate work. However, autonomous quantum heat engines have not yet been experimentally demonstrated in any system despite numerous theoretical investigations. Here, we experimentally realize an autonomous quantum heat engine based on superconducting circuits. We construct the engine circuit implementing an approximate Otto cycle by coupling two superconducting resonators with a superconducting quantum interference device, and coupling this system to spectrally filtered hot and cold reservoirs. By varying the experimental conditions, we observe coherent microwave power generation arising from the internal dynamics of the system driven only by the thermal reservoirs. Our results validate previous theoretical predictions for this circuit and pave the way for detailed studies of quantum effects in heat engines and for using heat-generated coherent microwaves in circuit quantum electrodynamics.
\end{abstract}

\maketitle

Heat engines are widely utilized devices that convert heat flow into usable energy or work. These devices have been extensively explored since the dawn of classical thermodynamics, and more recently, in the context of quantum thermodynamics with the use of meso- and microscopic working media carrying few excitation quanta in the operating regime~\cite{gemmer_quantum_2009,kosloff_quantum_2013,deffner_quantum_2019}. Experimental investigations of quantum heat engines~(QHE) may reveal the significance of quantum phenomena in their operation and potentially expand the toolbox of energy manipulation in small-scale quantum devices. Driven quantum heat engines have been realized in multiple physical systems such as a single-particle spin coupled to single-ion motion~\cite{von_lindenfels_spin_2019,van_horne_single-atom_2020}, a nuclear spin system with nuclear magnetic resonance~\cite{peterson_experimental_2019,de_assis_efficiency_2019}, nitrogen-vacancy center in a diamond~\cite{klatzow_experimental_2019}, QHE driven by atomic collisions of cesium and rubidium~\cite{bouton_quantum_2021}, and recently in superconducting circuits using a transmon qubit with quantum-circuit refrigeration~\cite{uusnakki_experimental_2025}.

During the past decade, superconducting circuits have attracted substantial interest due to their central role in numerous quantum-technological applications such as quantum computing~\cite{dicarlo_demonstration_2009,harrigan_quantum_2021,google_quantum_ai_and_collaborators_quantum_2025}, communication~\cite{kurpiers_deterministic_2018,axline_-demand_2018,fedorov_experimental_2021}, and sensing~\cite{gasparinetti_fast_2015,kokkoniemi_bolometer_2020,wang_quantum_2021}. Since these circuits are experimentally accessible and highly controllable, they constitute a versatile platform for investigating a variety of different quantum systems. Quantum-heat-engine realizations with superconducting circuits have been under investigation in several theoretical studies with resonators~\cite{niskanen_information_2007,campisi_nonequilibrium_2015,altintas_rabi_2015,karimi_otto_2016,marchegiani_self-oscillating_2016}, optomechanical systems~\cite{hardal_quantum_2017,rasola_proposal_2025}, and recently also in experimental studies~\cite{sundelin_quantum_2026,uusnakki_experimental_2025}. Quantum heat engines have also been extensively studied in other optomechanical configurations~\cite{zhang_quantum_2014,zhang_theory_2014,dong_work_2015,naseem_quantum_2019,izadyari_quantum_2022}. Other implementations of engineered thermal environments in superconducting circuits have also attracted persistent attention, including quantum-heat-transport experiments~\cite{pekola_towards_2015,ronzani_tunable_2018}, quantum measurement engines~\cite{elouard_extracting_2017,buffoni_quantum_2019,dassonneville_directly_2025}, quantum bath engineering~\cite{murch_cavity-assisted_2012,kimchi-schwartz_stabilizing_2016,sharafiev_leveraging_2025}, and local cooling of superconducting circuits using a quantum-circuit refrigerator~\cite{tan_quantum-circuit_2017,yoshioka_fast_2021,sevriuk_initial_2022,viitanen_quantum-circuit_2024,kivijarvi_noise-induced_2024}. 

Although quantum thermal machines have been experimentally realized on multiple platforms, all realizations thus far have required persistent driving and external control. This increases the difficulty of generating and observing more work output than the energy consumed by the external control field. However, autonomous quantum heat engines provide a platform to reduce the effort needed for external control, as their operation, in principle, only requires the flow of heat through the system to power the engine cycle. Although autonomous quantum heat engines have been studied in several theoretical works~\cite{tonner_autonomous_2005,kosloff_quantum_2014, roulet_autonomous_2018, niedenzu_concepts_2019,verteletsky_revealing_2020,rasola_autonomous_2024,rasola_proposal_2025}, their experimental investigations have proved challenging.

In this work, we demonstrate an autonomous quantum heat engine with a sinusoidally-modulated quantum Otto cycle using superconducting circuits based on a previous theoretical proposal~\cite{rasola_proposal_2025}. The working body of the quantum heat engine consists of a high-frequency superconducting resonator, coupled through a superconducting quantum interference device (SQUID) to a low-frequency driving resonator, such that the current flowing in the driving resonator essentially modulates the frequency of the working-body resonator. The hot and cold reservoirs of the device are implemented using two filtering resonators with Lorentzian-shaped spectra, which are coupled to dissipative thermal environments. To initialize the device operating point, we apply a constant magnetic flux through the SQUID loop, terminate the cold filter resonator at the base temperature of the cryostat, and apply a constant thermal-noise source to the drive line of the hot filter resonator. 

After calibrating the setup, we measure the transmission of a microwave signal through a feedline coupled to the driving resonator for different probe and noise powers. At certain noise and probe powers, we observe a shift and sharpening of the resonance dip of the driving resonator corresponding to an increased quality factor of the driving resonator owing to the input noise. We carry out ring-down experiments by introducing an initialization pulse through the feedline and measuring the radiation emitted from the driving resonator after varying delays. At large enough noise powers, we observe a major increase in the output power that saturates over time. To verify autonomous photon generation, we measure the bare output-signal from the driving resonator without any coherent excitation and observe a Lorentzian-shaped power spectrum centered at the resonator frequency while only injecting thermal noise at the hot filter port. Obtaining histograms from the peak of this spectrum yields a distribution with Poissonian characteristics for the signal power and a ring in the signal in-phase (I) and quadrature (Q) plane. These results indicate the generation of coherent microwave power solely by thermal noise and verify the autonomous operation of the quantum heat engine.

\subsection*{Autonomous operation principle}

\begin{figure*}[t]
  \begin{center}
    \ifincludefigures\includegraphics[width=1\linewidth
    ]{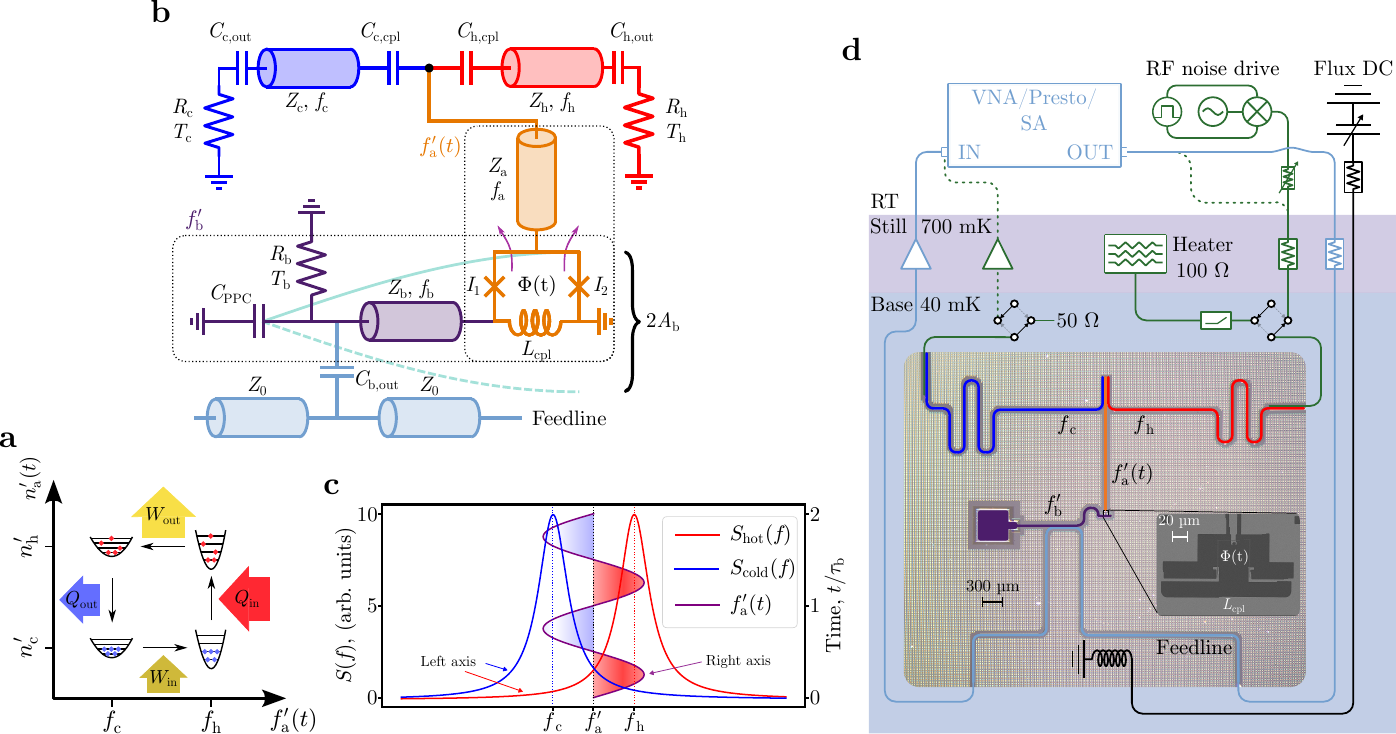}\fi     
  \end{center}
  \caption{\small\label{fig:setup}\textbf{Working principle and experimental setup of the autonomous quantum heat engine.} ~\textbf{a},~Illustration of an ideal quantum Otto cycle. The quantum working body is depicted as a quadratic potential energy in an eigenfrequency--photon-number plane. The red and blue arrows illustrate the heat coming in and out of the working body, respectively, and the yellow arrows similarly denote the related work. \textbf{b},~Circuit-element model of the autonomous quantum heat engine, comprised of five subcircuits: cold filter resonator + resistor (dark blue, c), hot filter resonator + resistor (red, h), working-body resonator + SQUID loop (orange, a), driving resonator (deep purple, b), and a probing feedline (light blue). For $x \in \{\rm a,b,c,h\}$, each resonator has impedance $Z_x$ and bare frequency $f_x$, and each resistor has resistance $R_x$ at temperature $T_x$. $C_{x,\rm out}$ and $C_{x,\rm cpl}$ denote output and coupling capacitors, $C_{\rm PPC}$ is the parallel plate capacitance, and $L_{\rm cpl}$ is the galvanic inductive coupling to the SQUID. The electric current with amplitude $A_\mathrm{b}$ of the driving resonator (cyan) modulates the effective flux $\Phi(t)$ through the SQUID with critical currents $I_1$ and $I_2$. Dashed boxes indicate the circuit components contributing to effective frequencies $f_\mathrm{a}'$ and $f_\mathrm{b}'$ of the working-body and driving resonators, respectively. \textbf{c},~Evolution of the working-body resonance frequency on top of the Lorentzian noise spectra of the hot and cold heat reservoirs as a function of frequency. \textbf{d},~Simplified experimental setup together with a false-color image of the heat engine circuit. The sample consists of hot (red) and cold (blue) filtering CPW resonators, the working-body resonator (orange), and the driving resonator (deep purple). The driving resonator is galvanically coupled to the working-body resonator through a SQUID (inset). A DC-driven magnetic coil controls the external flux outside the sample holder. We probe the driving resonator using a capacitively coupled feedline connected to selected measurement instruments at room temperature. For thermal-noise injection, we use both a 100-$\Omega$ heated resistor and an RF noise drive connected via a switch on the hot filter feedline.}
\end{figure*}

% Theory, principle of operation
The principle of operation for the autonomous quantum heat engine in this work is based on a non-linear coupling between two superconducting resonators. We aim to implement an approximate quantum Otto cycle in which heat and work strokes are executed continuously by modulating the occupation number and mode frequency of the working-body resonator. By considering the case of a single bosonic mode, the resulting change in internal energy can be described by the first law of thermodynamics as $\mathrm{d}E = n \mathrm{d}\mathcal{E} + \mathcal{E} \mathrm{d}n = \delta W + \delta Q$, where $n$ is the average photon number of the bosonic mode with energy quantum $\mathcal{E}$, $W$ is work, and $Q$ is heat. The individual strokes of the quantum Otto cycle are illustrated in Fig.~\ref{fig:setup}a. In short, we aim to extract additional work from the cycle by matching energy exchange with autonomous frequency modulation: the photon number increases at a high mode frequency and decreases at a low mode frequency, leading to positive net work over the cycle.

% Frequency tuning
To tune the mode frequency of our quantum working body, we couple a low-frequency driving resonator to the high-frequency working-body resonator using a SQUID such that a segment of the inductive part of the driving resonator is galvanically coupled to the superconducting loop of the SQUID. In this construction, the magnetic flux generated by the oscillating current in the driving resonator threads the SQUID loop and thus periodically modulates the frequency of the high-frequency resonator. This is displayed schematically in the lumped-element circuit diagram in Fig.~\ref{fig:setup}b. We can describe the coupled resonators with the approximate Hamiltonian~\cite{johansson_optomechanical-like_2014}
\begin{equation}
    \hat{H}(t) = hf_\mathrm{a}'[\Phi(t)]\hat{a}^\dagger\hat{a} + hf_\mathrm{b}'\hat{b}^\dagger\hat{b} - h g_0[\Phi(t)] \hat{a}^\dagger\hat{a}\left(\hat{b} + \hat{b}^\dagger\right),
\end{equation}
where $\hat{a}$ and $\hat{a}^\dagger$ are the annihilation and creation operators of the working-body resonator, respectively, $\hat{b}$ and $\hat{b}^\dagger$ are the annihilation and creation operators of the driving resonator, respectively, $f_\mathrm{a}'$ is the SQUID-modified frequency of the working-body resonator, $f_\mathrm{b}'$ is the effective frequency of the driving resonator, $\Phi(t) =~ \Phi_\mathrm{ext} + \Phi_\mathrm{b}(t)$ is the flux threading the SQUID loop comprised of an external flux and the flux induced by the driving resonator, and $g_0$ is a flux-dependent coupling constant.

% Heat reservoirs and coupling
For the heat reservoirs, we introduce a hot and a cold resistor with temperatures $T_\mathrm{h}$ and $T_\mathrm{c}$, respectively, each coupled to their respective filter resonator, which are capacitively coupled to the working-body resonator, as depicted in Fig.~\ref{fig:setup}b. The filter resonators are designed to have broad Lorentzian spectra with distinct central frequencies, thereby suppressing resonant coupling, while keeping them within the tunability range of the working-body resonator. The working principle of the device is elaborated in Fig.~\ref{fig:setup}c, where we illustrate the modulation of the resonance frequency of the working-body resonator over the period of the driving resonator, as well as the coupling to the filters with Lorentzian power spectra. The slow resonator will periodically drive the resonance frequency of the working-body resonator between the central frequencies of the power spectra of the reservoirs, thus periodically coupling strongly to the hot and cold reservoirs~\cite{rasola_proposal_2025}. In this fashion, the eigenfrequency of the working body is 
periodically tuned up and down while the system gains energy from the hot reservoir and releases it to the cold reservoir.

Since the filter spectra have infinite tails, the coupling to each reservoir is never completely extinguished, and hence, instead of having ideal adiabatic and isochoric strokes for the Otto cycle, the cycle of the heat engine is only an approximate sinusoidally-modulated Otto cycle. Examples of such cycles depicted in the occupation-number--eigenfrequency plane are displayed in Ref.~\cite{rasola_proposal_2025}. In a classical sense, the frequency modulation of the working-body resonator by the state of the driving resonator is proportional to the amplitude $A_\mathrm{b}$ of the field in the driving resonator. Therefore, in the engine cycle, the dynamic ranges of both the effective occupation number $n'_\mathrm{a}(t)$ and the eigenfrequency $f'_\mathrm{a}(t)$ increase as a function of this amplitude~\cite{rasola_proposal_2025}.

% Microwave power from thermal noise
With adequate parameters and initial conditions, the Otto cycle driven by the driving resonator and the temperature gradient between the heat reservoirs will result in net microwave power generation stored in the state of the driving resonator. Under such conditions, the state of the driving resonator will eventually reach a stable point of operation with maximal power output depending on the temperature gradient between the thermal noise sources $T_\mathrm{h}$ and $T_\mathrm{c}$, as well as the intrinsic dissipation rate of the resonator $\gamma_\mathrm{b}$~\cite{rasola_proposal_2025}.

\subsection*{Heat engine device and characterization}

In our experiment, the electrical circuit of the autonomous quantum heat engine consists of three coplanar-waveguide resonators (CPW), a tadpole resonator, and their respective coupling elements~\cite{blais_cavity_2004,goppl_coplanar_2008,rasola_low-characteristic-impedance_2024}. The microwave probe and thermal-noise circuitry, along with the sample chip, are displayed in Fig.~\ref{fig:setup}d, and a more detailed measurement setup is shown in Fig.~\ref{fig:ext_measurement_wiring}. The working-body resonator is galvanically contacted to the driving resonator with a SQUID loop to form the desired coupling between the states of the two resonators~\cite{johansson_optomechanical-like_2014,rasola_proposal_2025}. 

The driving resonator is realized by shunting a short strip of CPW with a large parallel-plate capacitor, resulting in a tadpole resonator design which is used to decrease the footprint of the resonator on the chip, to enable strong inductive coupling, and to avoid higher modes near the fundamental mode or the frequencies of the other resonators in the device ~\cite{rasola_low-characteristic-impedance_2024}. The working-body resonator is capacitively coupled to two filtering resonators to provide thermal-reservoir coupling. Both filtering resonators have capacitively coupled transmission lines, which are used both to characterize the resonators and to inject thermal noise. The broad Lorentzian shapes of the filter spectra are achieved by strong coupling to their respective drive lines, leading to low quality factors required for efficient heat engine operation~\cite{rasola_proposal_2025}. 

\begin{figure*}[t]
  \begin{center}
    \ifincludefigures\includegraphics[width=1\linewidth
    ]{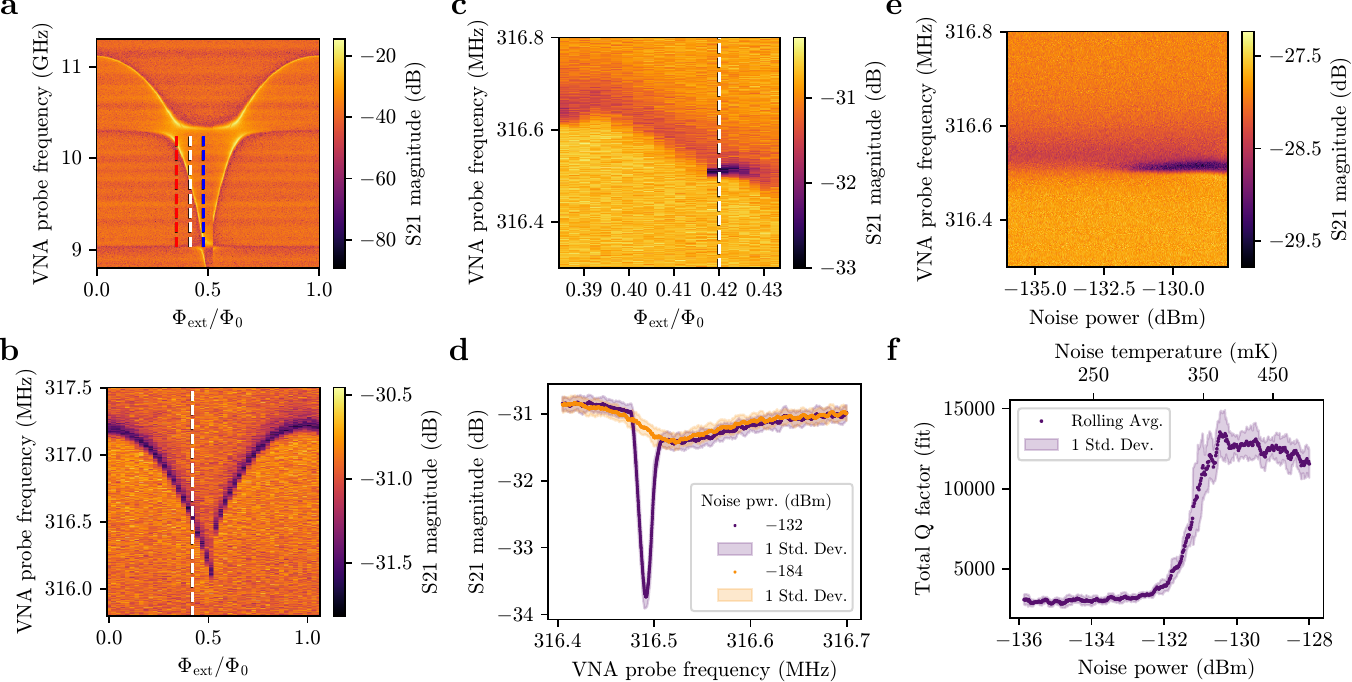}\fi     
  \end{center}
  \caption{\small\label{fig:characterization}\textbf{Engine characterization and calibration.}~\textbf{a, b},~Magnitude of the microwave transmission coefficient as a function of normalized external flux and probe frequency for (\textbf{a}) the filtering and working-body resonators and (\textbf{b}) the driving resonator. The red and blue dashed lines in (\textbf{a}) depict the avoided crossings between the working-body resonator and the filtering resonators with matching colors to Fig.~\ref{fig:setup}d. The optimal flux point for the heat engine is marked with white dashed lines in (\textbf{a}) and (\textbf{b}). \textbf{c},~As (\textbf{b}) but measured in the vicinity of the optimal flux point with RF noise on at $-132$\;dBm power and $-125$\;dBm probe power. The white dashed line displays the flux value for which the resonance dip of the driving resonator is most enhanced. \textbf{d},~Trace of the transmission magnitude along the white dashed line in (\textbf{c}) for RF noise on and off. \textbf{e},~Transmission magnitude as a function of the RF noise power and probe frequency at the optimal flux point for $-129$\;dBm probe power. \textbf{f},~Total quality factor of the driving resonator extracted from (\textbf{e}) as a function of the RF noise power (bottom axis) and calculated noise temperature (top axis). The rolling average window size is 10 samples. The default parameters of the device are listed in Table~\ref{tab:exp_parameters}.}
\end{figure*}

We characterize the resonators by measuring transmission spectroscopies while sweeping the external flux through the SQUID with a magnetic coil. For the filter and working-body resonators, we probe through the drive line coupled to the hot filter and measure the transmission through the filters and the working-body mode at the output of the cold filter. To probe the driving resonator, an additional capacitively coupled feedline is used. The results of the transmission spectroscopies are displayed in Fig.~\ref{fig:characterization}a for the filtering and working-body resonators, and Fig.~\ref{fig:characterization}b for the driving resonator. From the results, we extract the resonance frequencies of the working-body resonator and the driving resonator modified by the SQUID, obtaining $f_\mathrm{a}' = 11.10\;\mathrm{GHz}$ and $f_\mathrm{b}' = 317.2\;\mathrm{MHz}$, respectively, whereas the resonance frequencies of the hot and cold filters at zero flux are $f_\mathrm{h} = 10.29\;\mathrm{GHz}$ and $f_\mathrm{c} = 9.035\;\mathrm{GHz}$, respectively. We extract the total quality factors of the resonators from the measured transmission coefficient and find for the driving resonator $Q_{\rm b} = 3000$, for the hot filtering resonator $Q_{\rm h} = 280$, and for the cold filtering resonator $Q_{\rm c} = 260$, which are comparable to the values used in the quasiclassical simulations of Ref.~\cite{rasola_proposal_2025}. The main device parameters are summarized in Table~\ref{tab:exp_parameters}.

\begin{table}[t]
    \caption{\textbf{Summary of the parameters of the measured device.} Here, $f_x$ is the resonance frequency of resonator $x\in\{\rm a,b,c,h\}$ for the given flux, $Q_x$ is the total quality factor of resonator $x$, $T_\mathrm{base}$ is the base temperature of the cryostat, and $g_{y\rm,cpl}$ is the coupling constant between the working body resonator and filtering resonator $y\in\{\rm h,c\}$. The other circuit parameters in Fig.~\ref{fig:setup}b are approximated from matching transmission spectroscopies with numerical simulations (Methods). These parameters are displayed in Table~\ref{tab:num_parameters}.}
    
    \begin{tabular}{cccc}
    \hline
        Parameter & Value & Parameter & Value\\
        \hline
        $f_\mathrm{h} (\Phi_\mathrm{ext}=0)$ & 10.29\;GHz& $Q_{\rm h}$ & 280\\
        $f_\mathrm{c} (\Phi_\mathrm{ext}=0)$ & 9.035\;GHz& $Q_{\rm c}$ & 260\\
        $f_\mathrm{b}' (\Phi_\mathrm{ext}=0)$ & 317.2\;MHz& $Q_{\rm b}$ & 3000\\
        $f_\mathrm{a}' (\Phi_\mathrm{ext}=0)$ & 11.10\;GHz & $T_\mathrm{base}$ & 40\;mK\\
        $f_\mathrm{b}' (\Phi_\mathrm{ext}=0.42\times\Phi_0)$ & 316.5\;MHz & $g_\mathrm{h,cpl}$ & 106\;MHz\\
        $f_\mathrm{a}' (\Phi_\mathrm{ext}=0.42\times\Phi_0)$ & 9.730\;GHz & $g_\mathrm{c,cpl}$ & 94.8\;MHz\\
        \hline
    \end{tabular}
    \label{tab:exp_parameters}
\end{table}

\subsection*{Experimental protocol and calibration}

% Protocol: how do we intend to measure an autonomous quantum heat engine
We aim to experimentally investigate the operation of the autonomous quantum heat engine by utilizing the working principle shown in Fig.~\ref{fig:setup}c. This is carried out in three steps: First, we configure the two constant heat reservoirs using the drive lines of the filtering resonators. We couple the cold filter to the thermal environment of the mixing chamber in the dilution refrigerator at 40\;mK, and inject thermal noise into the hot filter either using a blackbody heater or artificial RF noise from room temperature. Second, we set the constant magnetic-flux environment of the sample chip such that the flux threading the SQUID loop tunes the working-body resonator to have its resonance frequency between those of the filtering resonators. Finally, we excite and probe the heat engine via a feedline coupled to the driving resonator by measuring the change in the transmitted microwave signal. After characterizing and calibrating the system to optimized parameters, we can probe the driving resonator to observe autonomous photon generation arising from the heat engine cycle and stored in the oscillating field of the driving resonator~\cite{rasola_proposal_2025,rasola_autonomous_2024}.

% Frequency tuning and optimal spot
As illustrated in the working principle of Fig.~\ref{fig:setup}c, it is desirable that the resonance frequency of the working-body resonator resides between the central frequencies of the filtering resonator spectrums. To set this stage, we use the external flux generated by a magnetic coil. From the flux sweep in Fig.~\ref{fig:characterization}a, we can identify two avoided crossings, where the working-body resonator frequency crosses the resonance frequencies of the hot and cold filters. Since the SQUID also affects the frequency of the driving resonator, we observe a similar flux tunability in Fig.~\ref{fig:characterization}b with a flux jump around the half flux point due to the bistability of the SQUID. Based on theoretical analysis~\cite{rasola_proposal_2025}, the optimal flux point lies roughly where the frequency of the working-body resonator is halfway between the avoided crossings, which we identify to be around $\Phi_\mathrm{ext}/\Phi_0 = 0.4$ from Fig.~\ref{fig:characterization}a. 

% Operational point and initial noise on/off
To identify the optimal operating point, we measure the transmission through the feedline and sweep the external flux around the predicted optimal value while injecting quasithermal RF noise with $-132$\;dBm power into the hot-filter drive line. The quasithermal noise is generated with an RF source and an AWG with a nearly flat spectrum centered on the resonance frequency of the hot filter resonator (Methods) and controlled by adjusting the power of the RF noise source. We observe a strong enhancement of the resonance dip of the driving resonator around the flux point $\Phi_\mathrm{ext}/\Phi_0 = 0.42$ in Fig.~\ref{fig:characterization}c. The difference between the resonance feature with and without injected RF noise is highlighted in Fig.~\ref{fig:characterization}d. When the noise is turned on, the depth of the resonance dip increases significantly, and the resonance feature shifts to a somewhat smaller frequency with a noticeably decreased linewidth. 

% Sweep thermal noise power, observe increase in total quality factor as a function of noise power
To quantify the change in the resonance dip, we sweep the noise power around the emergence point of the enhancement with a low probe power in Fig.~\ref{fig:characterization}e, and extract the total quality factor displayed as a rolling average in Fig.~\ref{fig:characterization}f. From these figures, we observe an increase in the total quality factor from approximately 3000 to 13000 in a smooth but step-like manner, centered at $-131$\;dBm of noise power. This roughly corresponds to a noise temperature of 350\;mK calculated from the RF noise power (Methods). Compared to the previous measurements in Fig.~\ref{fig:characterization}c with higher probe power and lower noise power, the decrease of probe power also shifts the resonance enhancement point to higher noise power, i.e., it appears that with a weaker excitation on the drive resonator, its possible extraction of work from the hot reservoir is smaller. 

\begin{figure*}[t]
  \begin{center}
    \ifincludefigures\includegraphics[width=1\linewidth
    ]{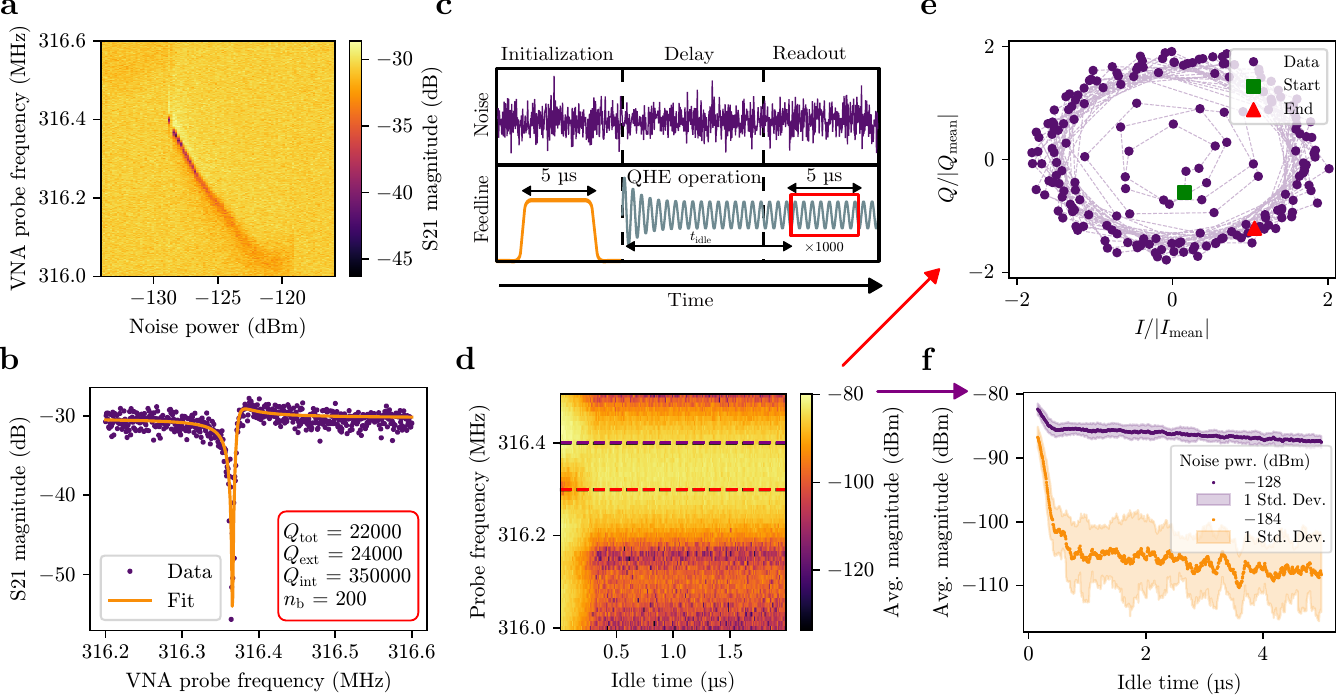}\fi     
  \end{center}
  \caption{\small\label{fig:RF_results}\textbf{Heat engine characteristics in transmission measurements.}~\textbf{a},~Transmission magnitude as a function of RF noise power and probe frequency at $-140$\;dBm probe power. \textbf{b},~Transmission magnitude as a function of probe frequency for $-150$\;dBm probe power and $-128$\;dBm noise power around the minimum point of (\textbf{a}). The obtained quality factors and the average photon number of the driving resonator from the fit are displayed in the inset. \textbf{c},~Measurement scheme for the time-domain measurements. The top panel illustrates the RF noise continuously fed through the drive line of the hot filter, and the bottom panel shows the kick pulse and the averaged integration window of the outgoing signal in the sample feedline. \textbf{d},~Average signal amplitude as a function of idle time and kick pulse frequency at $-128$\;dBm noise power, measured using the protocol in (\textbf{c}). The red dashed line marks the dip of the resonance feature at 316.3\;MHz probed in (\textbf{e}), and the purple dashed line marks the peak of the resonance feature at 316.4\;MHz probed in (\textbf{f}). \textbf{e},~Evolution of the resonator state as a function of idle time in the normalized IQ plane sliced from the purple dashed line in (\textbf{d}). The lines between points highlight the direction of temporal evolution with 10\;ns idle-time steps between points. \textbf{f},~Decay of the resonator state excitation as average signal amplitude as a function of the idle time for noise on and off for up to 5\;\textmu s evolution. The rolling average window is 30 samples. The default parameters of the device are listed in Table~\ref{tab:exp_parameters}.}
\end{figure*}

% Higher noise power and lower probe -> very high Q_int and low photon count
To further test the effect of the probe power and, consequently, the number of photons in the driving resonator, we probe the system with an even lower probe power and again sweep the noise power. The results are shown in Fig.~\ref{fig:RF_results}a, where we observe the emergence of the enhancement being pushed to even higher noise powers. We also observe a downward shift in the driving-resonator frequency with increasing noise temperature, as well as a much deeper dip. For a probe power providing roughly 200 photons to the driving resonator and using the noise power at the minimum point in Fig.~\ref{fig:RF_results}a, we observe an even deeper resonance dip of more than 25\;dB displayed in Fig.~\ref{fig:RF_results}b. Fitting this resonance feature yields a drastically improved internal quality factor of $Q_\mathrm{int} = 350\,000$, which is two orders of magnitude higher than the bare value, $Q_\mathrm{int}\sim 3000$, extracted from the bare resonance feature without noise.

This indicates that, even for a relatively low photon population in the driving resonator, the internal dissipation rate approaches zero, so that the external coupling limits the total dissipation. This behavior agrees with previous theoretical investigations of the circuit, where the internal dissipation of the driving resonator crosses zero into negative dissipation (photon generation) when the temperature of the hot reservoir is increased~\cite{rasola_proposal_2025}. However, the theoretical description does not account for the effects of the probe feedline and the resulting destructive and constructive interference of microwave photons. The quality factor of the resonator increases drastically with less probe power, which is the opposite for a regular tadpole resonator~\cite{rasola_low-characteristic-impedance_2024}, which might indicate that higher power disturbs the autonomous photon generation with destructive interference.

\subsection*{Autonomous photon generation}

%% Figure 4: Display spectrum analyzer results and distribution %%
\begin{figure*}[t]
  \begin{center}
    \ifincludefigures\includegraphics[width=1\linewidth
    ]{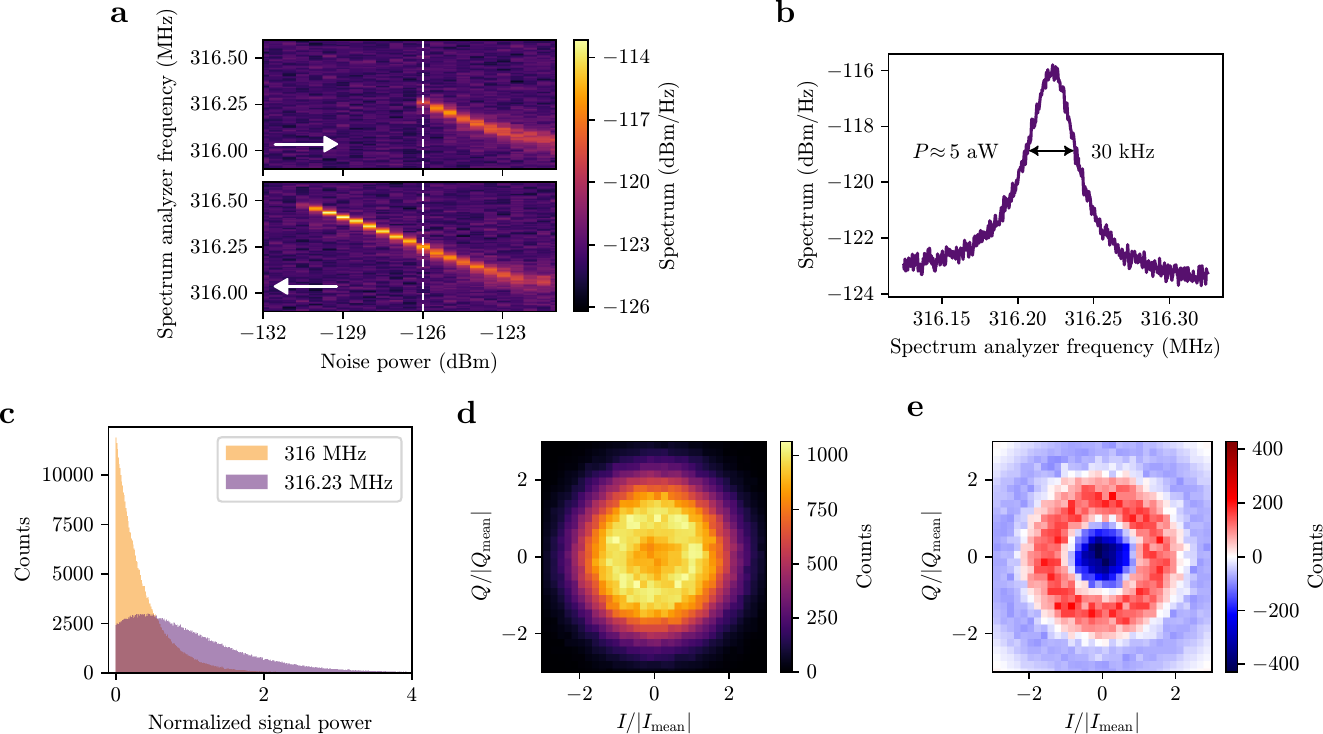}\fi     
  \end{center}
  \caption{\small\label{fig:spectrum_results}\textbf{Autonomous photon generation and power distribution.}~\textbf{a},~Power spectrum as a function of RF noise power and spectrum analyzer frequency. The direction of the noise power sweep is from low to high in the top panel and high to low in the bottom panel, indicated by the white arrows. The white dashed line marks the spontaneous emergence point of the emission. \textbf{b},~Power spectrum as a function of spectrum analyzer frequency for $-126$\;dBm noise power. The linewidth of the spectrum is 30\;kHz, and the integrated power from the spectrum is $P\approx 5$ aW. \textbf{c},~Histogram of normalized single shots from the incoming signal power of the heat engine for on and off peak of the spectrum in (\textbf{a}) for $-126$\;dBm noise power. \textbf{d},~Distribution of single shots from the signal generated by the heat engine in the normalized IQ plane for $-126$\;dBm noise power. \textbf{e},~Difference between the distribution in (\textbf{d}) and a Gaussian fitted to that distribution in the normalized IQ plane. The length of the integration windows for (\textbf{c}), (\textbf{d}), and (\textbf{e}) is 20\;\textmu s. The default parameters of the device are listed in Table~\ref{tab:exp_parameters}.}
\end{figure*}

% Time domain: State is sustained as a function of time
To more closely examine the behavior of the resonator around its resonance frequency in the high-noise, low-probe-power regime, we carry out time-domain spectroscopy at the optimal point indicated in Fig.~\ref{fig:RF_results}a. This is conducted using a decay-measurement protocol shown in Fig.~\ref{fig:RF_results}c, where we inject constant RF noise into the hot filter, send a 5\;\textmu s initialization/kick pulse at the desired frequency through the probe line, wait for an idle time of length $t_\mathrm{idle}$, and measure the amplitude and phase of the incoming signal from the probe line by time-integrating the signal over 5\;\textmu s. This is repeated 1000 times for each frequency and idle time value. 

The results of the time-domain spectroscopy are displayed in Fig.~\ref{fig:RF_results}d around the resonance feature. The slice at zero idle time roughly corresponds to the data in Fig.~\ref{fig:RF_results}b. We observe from the spectroscopy that the excitation in the driving resonator, initialized by the kick pulse, is sustained for frequencies around the resonance dip with a linewidth of about 300\;kHz, but quickly decays for frequencies further away from the resonance frequency. Exactly at resonance, the signal amplitude briefly increases before saturating to a stable value. The evolution of the signal at the resonance frequency is mapped into the IQ plane in Fig.~\ref{fig:RF_results}e, which depicts how the amplitude of the signal increases from the start and coherently evolves for the duration of the sweep instead of decaying into Gaussian noise. The difference between the regular decay of the resonator and the nearly sustained excitation at longer idle times is illustrated in Fig.~\ref{fig:RF_results}f, which shows the signal amplitude decay over 5\;\textmu s with the noise source turned on and off with kick pulse frequency at 316.4\;MHz. From the rolling averages and uncertainty estimates, we observe that the signal decays to the noise floor of the system within 1\;\textmu s when the noise is off, whereas the decay is much slower with noise on. These results indicate that when we inject quasithermal RF noise into the hot filter, the coherent state of the driving resonator is sustained for kick pulse frequencies around its resonance frequency. This corroborates the observed noise-induced improvement of the quality factor and shows that the lifetime of the resonator excitation is increased drastically, which may indicate autonomous photon generation arising from the heat engine cycle.

%% Figure 5: Display flux sweep, and temperature results from VNA and spectrum analyzer with heater as a noise source %%
\begin{figure*}[t]
  \begin{center}
    \ifincludefigures\includegraphics[width=1\linewidth
    ]{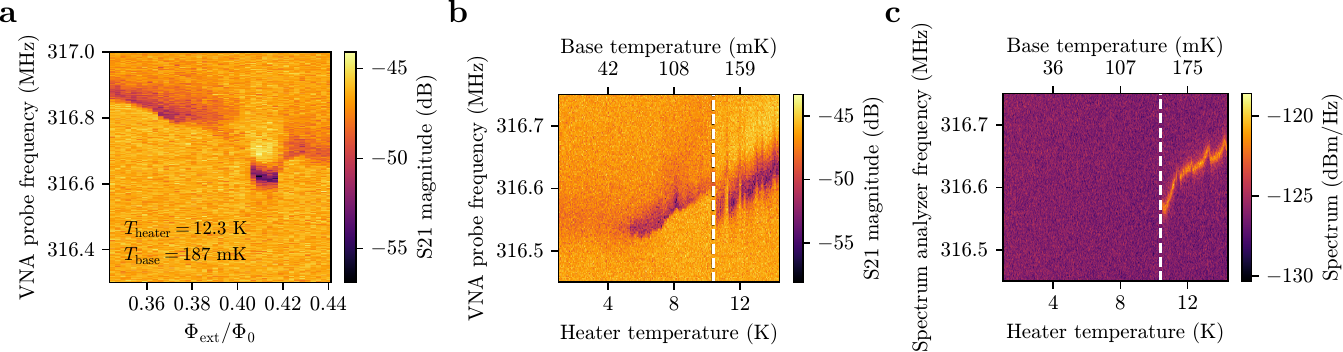}\fi     
  \end{center}
  \caption{\small\label{fig:heater_results}\textbf{Blackbody heater as a thermal noise source.}~\textbf{a},~Transmission magnitude as a function of normalized external flux and probe frequency with blackbody heater on. The temperature of the blackbody heater is approximately 12.3\;K, and the base temperature is 187\;mK. \textbf{b},~Transmission magnitude as a function of heater temperature (bottom axis), base temperature (top axis), and probe frequency. \textbf{c},~Power spectrum as a function of heater temperature (bottom axis), base temperature (top axis), and spectrum analyzer frequency. The temperature for the emergence point of the emission is marked with white dashed lines in (\textbf{b}) and (\textbf{c}). The default parameters of the device are listed in Table~\ref{tab:exp_parameters}.}
\end{figure*}

% Spectrum analyzer: Visible microwave generation peak with no probe tone from feedline
To verify that the time-domain measurements truly indicate photon generation, we measure the spectrum of the bare output signal coming from the feedline coupled to the driving resonator using a spectrum analyzer without any coherent excitation of the driving resonator. We measure the spectrum around the frequency range observed in Fig.~\ref{fig:RF_results}d and sweep the RF noise power first from low to high power. From the results shown in Fig.~\ref{fig:spectrum_results}a, we observe an emission peak emerging in a stepwise manner at $-126$\;dBm noise power. The emission peak shifts in frequency similarly to Fig.~\ref{fig:RF_results}a as the noise power increases. This indicates that at a strong enough noise, the autonomous photon generation from the heat engine cycle becomes greater than the intrinsic dissipation rate of the driving resonator, resulting in the photon generation observed as this emission peak. The averaged spectrum at the emergence power is shown in Fig.~\ref{fig:spectrum_results}b, which displays a Lorentzian-shaped emission peak with 30\;kHz linewidth. Integrating over the spectrum and taking into account the approximate gain of 75\;dBm from the amplifiers in the signal output line, we obtain an output power of $P \approx 5$~aW for the heat engine, which is in a similar power range as in the theoretical work~\cite{rasola_proposal_2025}. Remarkably, this emission feature is stable for hours, which is significantly longer than the period $\tau_\mathrm{b}\approx3.2$\;ns of the heat engine.

% Directional dependence: If the photon count is higher due to higher noise power, the photon generation is sustained longer -> matches transmission measurements with VNA at low power/high noise
In Fig.~\ref{fig:spectrum_results}a, we also sweep the noise power in the opposite direction, from high to low power. In contrast to the other direction, the emission peak is visible for lower noise powers. This hysteresis implies that once the heat engine has been started by the initially stronger noise, it can also run at a weaker noise. In addition, the linewidth of the spectrum further decreases for the lower noise powers down to around 15\;kHz. This is comparable to the transmission measurements, where we observed an increase in the quality factor and sustained excitation in the time-domain measurements for these lower noise temperatures. This indicates that for the heat engine cycle to start and run at low noise powers, the photon number in the driving resonator has to be high enough for the SQUID-mediated tuning of the working-body resonator to be large enough, while simultaneously the noise temperature in the hot filter has to be high enough to provide enough photons for the working-body resonator. In the case of the spontaneously emerging emission peak at $-126$\;dBm, the temperature of the hot filter is high enough such that the natural fluctuations in the driving resonator are strong enough to initiate the heat engine cycle and achieve stable operation, which is consistent with the theory~\cite{rasola_proposal_2025}.

% Distributions: Even though the emission spectrum is weak, the power distribution (Poissonian) differs from the Gaussian noise background. This is also visible in 2D histogram vs Gaussian fit, as the ring/donut in the IQ plane emerges 
We further investigate the spectral properties and phase coherence of the emission by measuring statistics of the received signal power and its quadratures. Since the emission is very weak and only $\sim$ 6\;dBm higher than the noise floor, the Gaussian distribution of noise photons in the sample environment plays a significant role, making it difficult to separate actual features of the generated photons from the noise. However, we can still identify signatures of coherent microwave generation from the distributions. In Fig.~\ref{fig:spectrum_results}c, we show the histograms of the normalized signal power of the bare signal coming from the sample through the feedline. The data are acquired from 400\,000 single-shot traces with 20\;\textmu s integration windows for two demodulation frequencies on and off the emission spectrum peak. The off-peak distribution for the 316 MHz demodulation frequency shows an exponential distribution peaked at zero power, in agreement with the distribution of a thermal field. However, the on-peak distribution at 316.23\;MHz displays Poissonian characteristics clearly different from the off-peak noise distribution. This indicates a coherent generation of photons in the incoming signal. 

In addition, we analyze the generated signal in the IQ plane in Fig.~\ref{fig:spectrum_results}d, where we have a two-dimensional (2D) histogram of the I and Q quadratures of the signal. As stated above, the effect of noise photons is significant, leading to a large number of points in the vicinity of the origin of the IQ plane. However, we can still clearly distinguish a larger number of points around the center in a ring corresponding to coherent photons. Due to noticeable drifts and fluctuations of the signal amplitude and phase, we could not investigate the time evolution of the signal in the same manner as in Fig.~\ref{fig:RF_results}e as we are lacking a reproducible signal starting point at which we know the phase of the signal. To further argue that the signal is not pure noise, we fit a 2D Gaussian distribution to that in Fig.~\ref{fig:spectrum_results}d and take the difference between the fit and the measured data. The result is displayed in Fig.~\ref{fig:spectrum_results}e, where we clearly observe that the measured distribution is not Gaussian and that the I and Q points integrated from the signal form a ring around zero. This indicates that the generated signal from the operation of the autonomous quantum heat engine is coherent with a small amplitude and drifting phase, partially masked by the noise floor in the incoming signal.

% Heater measurements: Most of the same observations (enhanced resonance, Q factor increase, sustained excitation, spectrum analyzer) are replicated with the actual heater
We replicate the previous measurements by switching away from the quasithermal RF noise source and using instead the blackbody heater described in detail in the Methods. We control the temperature of the heater with a PID controller and monitor its temperature with a thermometer connected to the heater. By increasing the temperature of the heater above 5\;K, we observe similar effects on the resonance of the driving resonator as with the RF noise. This is depicted in Fig.~\ref{fig:heater_results}a, where we display a flux spectroscopy over the operation point similar to Fig.~\ref{fig:characterization}c. The resonance dip deepens, and we observe a similar downward shift in the resonance frequency as observed for the high RF noise power in Fig.~\ref{fig:RF_results}a. We measured changes in the resonance spectrum as a function of the heater temperature by recording the spectrum every second, while manually adjusting and logging the heater temperature at 10-second intervals using the cryostat control software. We also record the temperature of the cryostat mixing chamber plate, where the sample holder is thermalized. Naturally, this temperature will estimate the sample temperature with a delay. We observe that the resonance frequency shifts higher when the temperature of the sample starts to increase, but the enhanced resonance feature is sustained. Finally, we measure the spectrum of the bare signal coming from the feedline and observe an emerging emission spectrum similar to Fig.~\ref{fig:spectrum_results}a after the heater temperature is increased to roughly 10.4\;K. These temperatures exceed the noise temperatures estimated in Fig.~\ref{fig:characterization}f, which can be attributed to the substantial insertion loss associated with the heater setup (Methods). The linewidth of the spectrum is smaller than in Fig.~\ref{fig:spectrum_results}b, with a mean value of 20\;kHz. These results indicate that the device can be operated using a natural radiating thermal source, and we can autonomously generate coherent microwaves purely by using thermal radiation.

\subsection*{Discussion}
In conclusion, we have experimentally realized an autonomous quantum heat engine with non-linear superconducting circuits using both artificial and natural thermal noise. We have demonstrated that we can observe coherent-microwave generation arising solely from the internal dynamics of the driving resonator and working-body system coupled to hot and cold thermal reservoirs. We have observed an emission spectrum that lasts for hours using a spectrum analyzer with an output power in the range of attowatts, agreeing with quasiclassically simulated values~\cite{rasola_proposal_2025}. This work provides the first proof-of-concept for autonomously operating a quantum heat engine without persistent external control or driving and demonstrates the versatility of different operational regimes.

Our findings confirm earlier theoretical treatments of autonomous quantum heat engines, improving our understanding of quantum thermodynamics and heat engines, and laying a foundation for future autonomous heat engine implementations and further research on thermal control of superconducting circuits. For example, it would be of great interest to realize this concept in other physical systems, such as in an optomechanical device. In the future, the performance of this device may be further enhanced and studied by improving the intrinsic quality factor of the driving resonator via better fabrication accuracy, which can lead to improved output power and efficiency of the engine~\cite{rasola_proposal_2025}. Such improvements may also allow to operate the engine in a regime with an enhanced role of quantum effects, and even study possibilities of extracting work from quantum correlation~\cite{perarnau-llobet_extractable_2015}.

%% References %%
\bibliographystyle{naturemag}
\bibliography{biblio}

@book{gemmer_quantum_2009,
	title = {Quantum thermodynamics: {Emergence} of thermodynamic behavior within composite quantum systems},
	volume = {784},
	publisher = {Springer},
	author = {Gemmer, Jochen and Michel, Mathias and Mahler, Günter},
	year = {2009},
}

@article{zhang_bridge-free_2017,
	title = {Bridge-free fabrication process for {Al}/{AlO}$_{\textrm{ \textit{x} }}$ /{Al} {Josephson} junctions},
	volume = {26},
	journal = {Chinese Physics B},
	author = {Zhang, Ke and Li, Meng-Meng and Liu, Qiang and Yu, Hai-Feng and Yu, Yang},
	year = {2017},
	pages = {078501},
}

@article{kosloff_quantum_2013,
	title = {Quantum thermodynamics: {A} dynamical viewpoint},
	volume = {15},
	journal = {Entropy},
	author = {Kosloff, Ronnie},
	year = {2013},
	pages = {2100--2128},
}

@book{deffner_quantum_2019,
	title = {Quantum {Thermodynamics}: {An} introduction to the thermodynamics of quantum information},
	publisher = {Morgan \& Claypool Publishers},
	author = {Deffner, Sebastian and Campbell, Steve},
	year = {2019},
}

@article{von_lindenfels_spin_2019,
	title = {Spin heat engine coupled to a harmonic-oscillator flywheel},
	volume = {123},
	journal = {Physical Review Letters},
	author = {Von Lindenfels, David and Gräb, Oliver and Schmiegelow, Christian T and Kaushal, Vidyut and Schulz, Jonas and Mitchison, Mark T and Goold, John and Schmidt-Kaler, Ferdinand and Poschinger, Ulrich G},
	year = {2019},
	pages = {080602},
}

@article{van_horne_single-atom_2020,
	title = {Single-atom energy-conversion device with a quantum load},
	volume = {6},
	journal = {npj Quantum Information},
	author = {Van Horne, Noah and Yum, Dahyun and Dutta, Tarun and Hänggi, Peter and Gong, Jiangbin and Poletti, Dario and Mukherjee, Manas},
	year = {2020},
	pages = {37},
}

@article{peterson_experimental_2019,
	title = {Experimental characterization of a spin quantum heat engine},
	volume = {123},
	journal = {Physical Review Letters},
	author = {Peterson, John PS and Batalhao, Tiago B and Herrera, Marcela and Souza, Alexandre M and Sarthour, Roberto S and Oliveira, Ivan S and Serra, Roberto M},
	year = {2019},
	pages = {240601},
}

@article{de_assis_efficiency_2019,
	title = {Efficiency of a {Quantum} {Otto} {Heat} {Engine} {Operating} under a {Reservoir} at {Effective} {Negative} {Temperatures}},
	volume = {122},
	journal = {Physical Review Letters},
	author = {De Assis, Rogério J. and De Mendonça, Taysa M. and Villas-Boas, Celso J. and De Souza, Alexandre M. and Sarthour, Roberto S. and Oliveira, Ivan S. and De Almeida, Norton G.},
	year = {2019},
	pages = {240602},
}

@article{klatzow_experimental_2019,
	title = {Experimental {Demonstration} of {Quantum} {Effects} in the {Operation} of {Microscopic} {Heat} {Engines}},
	volume = {122},
	journal = {Physical Review Letters},
	author = {Klatzow, James and Becker, Jonas N. and Ledingham, Patrick M. and Weinzetl, Christian and Kaczmarek, Krzysztof T. and Saunders, Dylan J. and Nunn, Joshua and Walmsley, Ian A. and Uzdin, Raam and Poem, Eilon},
	year = {2019},
	pages = {110601},
}

@article{bouton_quantum_2021,
	title = {A quantum heat engine driven by atomic collisions},
	volume = {12},
	journal = {Nature Communications},
	author = {Bouton, Quentin and Nettersheim, Jens and Burgardt, Sabrina and Adam, Daniel and Lutz, Eric and Widera, Artur},
	year = {2021},
	pages = {2063},
}

@misc{uusnakki_experimental_2025,
	title = {Experimental realization of a quantum heat engine based on dissipation-engineered superconducting circuits},
	author = {Uusnäkki, Tuomas and Mörstedt, Timm and Teixeira, Wallace and Rasola, Miika and Möttönen, Mikko},
	year = {2025},
	note = {arXiv:2502.20143},
}

@article{dicarlo_demonstration_2009,
	title = {Demonstration of two-qubit algorithms with a superconducting quantum processor},
	volume = {460},
	journal = {Nature},
	author = {DiCarlo, Leonardo and Chow, Jerry M and Gambetta, Jay M and Bishop, Lev S and Johnson, Blake R and Schuster, DI and Majer, J and Blais, Alexandre and Frunzio, Luigi and Girvin, SM and {others}},
	year = {2009},
	pages = {240--244},
}

@article{harrigan_quantum_2021,
	title = {Quantum approximate optimization of non-planar graph problems on a planar superconducting processor},
	volume = {17},
	journal = {Nature Physics},
	author = {Harrigan, Matthew P and Sung, Kevin J and Neeley, Matthew and Satzinger, Kevin J and Arute, Frank and Arya, Kunal and Atalaya, Juan and Bardin, Joseph C and Barends, Rami and Boixo, Sergio and {others}},
	year = {2021},
	pages = {332--336},
}

@article{google_quantum_ai_and_collaborators_quantum_2025,
	title = {Quantum error correction below the surface code threshold},
	volume = {638},
	journal = {Nature},
	author = {{Google Quantum AI and Collaborators} and Acharya, Rajeev and Abanin, Dmitry A. and Aghababaie-Beni, Laleh and Aleiner, Igor and Andersen, Trond I. and Ansmann, Markus and Arute, Frank and Arya, Kunal and Asfaw, Abraham and Astrakhantsev, Nikita and Atalaya, Juan and Babbush, Ryan and Bacon, Dave and Ballard, Brian and Bardin, Joseph C. and Bausch, Johannes and Bengtsson, Andreas and Bilmes, Alexander and Blackwell, Sam and Boixo, Sergio and Bortoli, Gina and Bourassa, Alexandre and Bovaird, Jenna and Brill, Leon and Broughton, Michael and Browne, David A. and Buchea, Brett and Buckley, Bob B. and Buell, David A. and Burger, Tim and Burkett, Brian and Bushnell, Nicholas and Cabrera, Anthony and Campero, Juan and Chang, Hung-Shen and Chen, Yu and Chen, Zijun and Chiaro, Ben and Chik, Desmond and Chou, Charina and Claes, Jahan and Cleland, Agnetta Y. and Cogan, Josh and Collins, Roberto and Conner, Paul and Courtney, William and Crook, Alexander L. and Curtin, Ben and Das, Sayan and Davies, Alex and De Lorenzo, Laura and Debroy, Dripto M. and Demura, Sean and Devoret, Michel and Di Paolo, Agustin and Donohoe, Paul and Drozdov, Ilya and Dunsworth, Andrew and Earle, Clint and Edlich, Thomas and Eickbusch, Alec and Elbag, Aviv Moshe and Elzouka, Mahmoud and Erickson, Catherine and Faoro, Lara and Farhi, Edward and Ferreira, Vinicius S. and Burgos, Leslie Flores and Forati, Ebrahim and Fowler, Austin G. and Foxen, Brooks and Ganjam, Suhas and Garcia, Gonzalo and Gasca, Robert and Genois, Élie and Giang, William and Gidney, Craig and Gilboa, Dar and Gosula, Raja and Dau, Alejandro Grajales and Graumann, Dietrich and Greene, Alex and Gross, Jonathan A. and Habegger, Steve and Hall, John and Hamilton, Michael C. and Hansen, Monica and Harrigan, Matthew P. and Harrington, Sean D. and Heras, Francisco J. H. and Heslin, Stephen and Heu, Paula and Higgott, Oscar and Hill, Gordon and Hilton, Jeremy and Holland, George and Hong, Sabrina and Huang, Hsin-Yuan and Huff, Ashley and Huggins, William J. and Ioffe, Lev B. and Isakov, Sergei V. and Iveland, Justin and Jeffrey, Evan and Jiang, Zhang and Jones, Cody and Jordan, Stephen and Joshi, Chaitali and Juhas, Pavol and Kafri, Dvir and Kang, Hui and Karamlou, Amir H. and Kechedzhi, Kostyantyn and Kelly, Julian and Khaire, Trupti and Khattar, Tanuj and Khezri, Mostafa and Kim, Seon and Klimov, Paul V. and Klots, Andrey R. and Kobrin, Bryce and Kohli, Pushmeet and Korotkov, Alexander N. and Kostritsa, Fedor and Kothari, Robin and Kozlovskii, Borislav and Kreikebaum, John Mark and Kurilovich, Vladislav D. and Lacroix, Nathan and Landhuis, David and Lange-Dei, Tiano and Langley, Brandon W. and Laptev, Pavel and Lau, Kim-Ming and Le Guevel, Loïck and Ledford, Justin and Lee, Joonho and Lee, Kenny and Lensky, Yuri D. and Leon, Shannon and Lester, Brian J. and Li, Wing Yan and Li, Yin and Lill, Alexander T. and Liu, Wayne and Livingston, William P. and Locharla, Aditya and Lucero, Erik and Lundahl, Daniel and Lunt, Aaron and Madhuk, Sid and Malone, Fionn D. and Maloney, Ashley and Mandrà, Salvatore and Manyika, James and Martin, Leigh S. and Martin, Orion and Martin, Steven and Maxfield, Cameron and McClean, Jarrod R. and McEwen, Matt and Meeks, Seneca and Megrant, Anthony and Mi, Xiao and Miao, Kevin C. and Mieszala, Amanda and Molavi, Reza and Molina, Sebastian and Montazeri, Shirin and Morvan, Alexis and Movassagh, Ramis and Mruczkiewicz, Wojciech and Naaman, Ofer and Neeley, Matthew and Neill, Charles and Nersisyan, Ani and Neven, Hartmut and Newman, Michael and Ng, Jiun How and Nguyen, Anthony and Nguyen, Murray and Ni, Chia-Hung and Niu, Murphy Yuezhen and O’Brien, Thomas E. and Oliver, William D. and Opremcak, Alex and Ottosson, Kristoffer and Petukhov, Andre and Pizzuto, Alex and Platt, John and Potter, Rebecca and Pritchard, Orion and Pryadko, Leonid P. and Quintana, Chris and Ramachandran, Ganesh and Reagor, Matthew J. and Redding, John and Rhodes, David M. and Roberts, Gabrielle and Rosenberg, Eliott and Rosenfeld, Emma and Roushan, Pedram and Rubin, Nicholas C. and Saei, Negar and Sank, Daniel and Sankaragomathi, Kannan and Satzinger, Kevin J. and Schurkus, Henry F. and Schuster, Christopher and Senior, Andrew W. and Shearn, Michael J. and Shorter, Aaron and Shutty, Noah and Shvarts, Vladimir and Singh, Shraddha and Sivak, Volodymyr and Skruzny, Jindra and Small, Spencer and Smelyanskiy, Vadim and Smith, W. Clarke and Somma, Rolando D. and Springer, Sofia and Sterling, George and Strain, Doug and Suchard, Jordan and Szasz, Aaron and Sztein, Alex and Thor, Douglas and Torres, Alfredo and Torunbalci, M. Mert and Vaishnav, Abeer and Vargas, Justin and Vdovichev, Sergey and Vidal, Guifre and Villalonga, Benjamin and Heidweiller, Catherine Vollgraff and Waltman, Steven and Wang, Shannon X. and Ware, Brayden and Weber, Kate and Weidel, Travis and White, Theodore and Wong, Kristi and Woo, Bryan W. K. and Xing, Cheng and Yao, Z. Jamie and Yeh, Ping and Ying, Bicheng and Yoo, Juhwan and Yosri, Noureldin and Young, Grayson and Zalcman, Adam and Zhang, Yaxing and Zhu, Ningfeng and Zobrist, Nicholas},
	year = {2025},
	pages = {920--926},
}

@article{kurpiers_deterministic_2018,
	title = {Deterministic quantum state transfer and remote entanglement using microwave photons},
	volume = {558},
	journal = {Nature},
	author = {Kurpiers, Philipp and Magnard, Paul and Walter, Theo and Royer, Baptiste and Pechal, Marek and Heinsoo, Johannes and Salathé, Yves and Akin, Abdulkadir and Storz, Simon and Besse, J-C and {others}},
	year = {2018},
	pages = {264--267},
}

@article{axline_-demand_2018,
	title = {On-demand quantum state transfer and entanglement between remote microwave cavity memories},
	volume = {14},
	journal = {Nature Physics},
	author = {Axline, Christopher J and Burkhart, Luke D and Pfaff, Wolfgang and Zhang, Mengzhen and Chou, Kevin and Campagne-Ibarcq, Philippe and Reinhold, Philip and Frunzio, Luigi and Girvin, SM and Jiang, Liang and {others}},
	year = {2018},
	pages = {705--710},
}

@article{fedorov_experimental_2021,
	title = {Experimental quantum teleportation of propagating microwaves},
	volume = {7},
	journal = {Science Advances},
	author = {Fedorov, Kirill G. and Renger, Michael and Pogorzalek, Stefan and Di Candia, Roberto and Chen, Qiming and Nojiri, Yuki and Inomata, Kunihiro and Nakamura, Yasunobu and Partanen, Matti and Marx, Achim and Gross, Rudolf and Deppe, Frank},
	year = {2021},
	pages = {eabk0891},
}

@article{gasparinetti_fast_2015,
	title = {Fast electron thermometry for ultrasensitive calorimetric detection},
	volume = {3},
	journal = {Physical Review Applied},
	author = {Gasparinetti, S and Viisanen, KL and Saira, O-P and Faivre, T and Arzeo, M and Meschke, Matthias and Pekola, Jukka P},
	year = {2015},
	pages = {014007},
}

@article{kokkoniemi_bolometer_2020,
	title = {Bolometer operating at the threshold for circuit quantum electrodynamics},
	volume = {586},
	journal = {Nature},
	author = {Kokkoniemi, Roope and Girard, J-P and Hazra, Dibyendu and Laitinen, Antti and Govenius, Joonas and Lake, RE and Sallinen, Iiro and Vesterinen, Visa and Partanen, Matti and Tan, JY and {others}},
	year = {2020},
	pages = {47--51},
}

@article{wang_quantum_2021,
	title = {Quantum {Microwave} {Radiometry} with a {Superconducting} {Qubit}},
	volume = {126},
	journal = {Physical Review Letters},
	author = {Wang, Zhixin and Xu, Mingrui and Han, Xu and Fu, Wei and Puri, Shruti and Girvin, S. M. and Tang, Hong X. and Shankar, S. and Devoret, M. H.},
	year = {2021},
	pages = {180501},
}

@article{niskanen_information_2007,
	title = {Information entropic superconducting microcooler},
	volume = {76},
	journal = {Physical Review B},
	author = {Niskanen, A. O. and Nakamura, Y. and Pekola, J. P.},
	year = {2007},
	pages = {174523},
}

@article{campisi_nonequilibrium_2015,
	title = {Nonequilibrium fluctuations in quantum heat engines: theory, example, and possible solid state experiments},
	volume = {17},
	journal = {New Journal of Physics},
	author = {Campisi, Michele and Pekola, Jukka and Fazio, Rosario},
	year = {2015},
	pages = {035012},
}

@article{altintas_rabi_2015,
	title = {Rabi model as a quantum coherent heat engine: {From} quantum biology to superconducting circuits},
	volume = {91},
	journal = {Physical Review A},
	author = {Altintas, Ferdi and Hardal, Ali {\"U}.C. and Müstecaplıoğlu, {\"O}zgür E.},
	year = {2015},
	pages = {023816},
}

@article{karimi_otto_2016,
	title = {Otto refrigerator based on a superconducting qubit: {Classical} and quantum performance},
	volume = {94},
	journal = {Physical Review B},
	author = {Karimi, B. and Pekola, J. P.},
	year = {2016},
	pages = {184503},
}

@article{marchegiani_self-oscillating_2016,
	title = {Self-oscillating {Josephson} quantum heat engine},
	volume = {6},
	journal = {Physical Review Applied},
	author = {Marchegiani, G and Virtanen, P and Giazotto, F and Campisi, M},
	year = {2016},
	pages = {054014},
}

@article{hardal_quantum_2017,
	title = {Quantum heat engine with coupled superconducting resonators},
	volume = {96},
	journal = {Physical Review E},
	author = {Hardal, Ali {\"U}.C. and Aslan, Nur and Wilson, C. M. and Müstecaplıoğlu, {\"O}zgür E.},
	year = {2017},
	pages = {062120},
}

@article{zhang_quantum_2014,
	title = {Quantum {Optomechanical} {Heat} {Engine}},
	volume = {112},
	journal = {Physical Review Letters},
	author = {Zhang, Keye and Bariani, Francesco and Meystre, Pierre},
	year = {2014},
	pages = {150602},
}

@article{zhang_theory_2014,
	title = {Theory of an optomechanical quantum heat engine},
	volume = {90},
	journal = {Physical Review A},
	author = {Zhang, Keye and Bariani, Francesco and Meystre, Pierre},
	year = {2014},
	pages = {023819},
}

@article{dong_work_2015,
	title = {Work measurement in an optomechanical quantum heat engine},
	volume = {92},
	journal = {Physical Review A},
	author = {Dong, Ying and Zhang, Keye and Bariani, Francesco and Meystre, Pierre},
	year = {2015},
	pages = {033854},
}

@article{naseem_quantum_2019,
	title = {Quantum heat engine with a quadratically coupled optomechanical system},
	volume = {36},
	journal = {Journal of the Optical Society of America B},
	author = {Naseem, M. Tahir and Müstecaplıoğlu, {\"O}zgür E.},
	year = {2019},
	pages = {3000},
}

@article{izadyari_quantum_2022,
	title = {Quantum signatures in a quadratic optomechanical heat engine with an atom in a tapered trap},
	volume = {39},
	journal = {Journal of the Optical Society of America B},
	author = {Izadyari, Mohsen and Öncü, Mehmet and Durak, Kadir and Müstecaplıoğlu, {\"O}zgür E.},
	year = {2022},
	pages = {3247},
}

@article{rasola_autonomous_2024,
	title = {Autonomous quantum heat engine based on non-{Markovian} dynamics of an optomechanical {Hamiltonian}},
	volume = {14},
	journal = {Scientific Reports},
	author = {Rasola, Miika and Möttönen, Mikko},
	year = {2024},
	pages = {9448},
}

@article{sundelin_quantum_2026,
	title = {Quantum refrigeration powered by noise in a superconducting circuit},
	volume = {17},
	journal = {Nature Communications},
	author = {Sundelin, Simon and Aamir, Mohammed Ali and Kulkarni, Vyom Manish and Castillo-Moreno, Claudia and Gasparinetti, Simone},
	year = {2026},
	pages = {359},
}

@article{pekola_towards_2015,
	title = {Towards quantum thermodynamics in electronic circuits},
	volume = {11},
	journal = {Nature Physics},
	author = {Pekola, Jukka P},
	year = {2015},
	pages = {118--123},
}

@article{ronzani_tunable_2018,
	title = {Tunable photonic heat transport in a quantum heat valve},
	volume = {14},
	journal = {Nature Physics},
	author = {Ronzani, Alberto and Karimi, Bayan and Senior, Jorden and Chang, Yu-Cheng and Peltonen, Joonas T and Chen, ChiiDong and Pekola, Jukka P},
	year = {2018},
	pages = {991--995},
}

@article{elouard_extracting_2017,
	title = {Extracting {Work} from {Quantum} {Measurement} in {Maxwell}’s {Demon} {Engines}},
	volume = {118},
	journal = {Physical Review Letters},
	author = {Elouard, Cyril and Herrera-Martí, David and Huard, Benjamin and Auffèves, Alexia},
	year = {2017},
	pages = {260603},
}

@article{buffoni_quantum_2019,
	title = {Quantum {Measurement} {Cooling}},
	volume = {122},
	journal = {Physical Review Letters},
	author = {Buffoni, Lorenzo and Solfanelli, Andrea and Verrucchi, Paola and Cuccoli, Alessandro and Campisi, Michele},
	year = {2019},
	pages = {070603},
}

@misc{dassonneville_directly_2025,
	title = {Directly probing work extraction from a single qubit engine fueled by quantum measurements},
	author = {Dassonneville, Rémy and Elouard, Cyril and Cazali, Romain and Assouly, Réouven and Bienfait, Audrey and Auffèves, Alexia and Huard, Benjamin},
	year = {2025},
	note = {arXiv:2501.17069},
}

@article{murch_cavity-assisted_2012,
	title = {Cavity-{Assisted} {Quantum} {Bath} {Engineering}},
	volume = {109},
	journal = {Physical Review Letters},
	author = {Murch, K. W. and Vool, U. and Zhou, D. and Weber, S. J. and Girvin, S. M. and Siddiqi, I.},
	year = {2012},
	pages = {183602},
}

@article{kimchi-schwartz_stabilizing_2016,
	title = {Stabilizing {Entanglement} via {Symmetry}-{Selective} {Bath} {Engineering} in {Superconducting} {Qubits}},
	volume = {116},
	journal = {Physical Review Letters},
	author = {Kimchi-Schwartz, M. E. and Martin, L. and Flurin, E. and Aron, C. and Kulkarni, M. and Tureci, H. E. and Siddiqi, I.},
	year = {2016},
	pages = {240503},
}

@article{sharafiev_leveraging_2025,
	title = {Leveraging {Collective} {Effects} for {Thermometry} in {Waveguide} {Quantum} {Electrodynamics}},
	volume = {134},
	journal = {Physical Review Letters},
	author = {Sharafiev, Aleksei and Juan, Mathieu and Cattaneo, Marco and Kirchmair, Gerhard},
	year = {2025},
	pages = {213602},
}

@article{tan_quantum-circuit_2017,
	title = {Quantum-circuit refrigerator},
	volume = {8},
	journal = {Nature Communications},
	author = {Tan, Kuan Yen and Partanen, Matti and Lake, Russell E and Govenius, Joonas and Masuda, Shumpei and Möttönen, Mikko},
	year = {2017},
	pages = {1--8},
}

@article{yoshioka_fast_2021,
	title = {Fast unconditional initialization for superconducting qubit and resonator using quantum-circuit refrigerator},
	volume = {119},
	journal = {Applied Physics Letters},
	author = {Yoshioka, T. and Tsai, J. S.},
	year = {2021},
	pages = {124003},
}

@article{sevriuk_initial_2022,
	title = {Initial experimental results on a superconducting-qubit reset based on photon-assisted quasiparticle tunneling},
	volume = {121},
	journal = {Applied Physics Letters},
	author = {Sevriuk, VA and Liu, W and Rönkkö, J and Hsu, H and Marxer, F and Mörstedt, TF and Partanen, M and Räbinä, J and Venkatesh, M and Hotari, J and {others}},
	year = {2022},
        pages = {234002},
}

@article{viitanen_quantum-circuit_2024,
	title = {Quantum-circuit refrigeration of a superconducting microwave resonator well below a single quantum},
	volume = {6},
	journal = {Physical Review Research},
	author = {Viitanen, Arto and Mörstedt, Timm and Teixeira, Wallace S and Tiiri, Maaria and Räbinä, Jukka and Silveri, Matti and Möttönen, Mikko},
	year = {2024},
	pages = {023262},
}

@misc{kivijarvi_noise-induced_2024,
	title = {Noise-induced quantum-circuit refrigeration},
	author = {Kivijärvi, Heidi and Viitanen, Arto and Mörstedt, Timm and Möttönen, Mikko},
	year = {2024},
	note = {arXiv:2412.05886},
}

@article{tonner_autonomous_2005,
	title = {Autonomous quantum thermodynamic machines},
	volume = {72},
	journal = {Physical Review E},
	author = {Tonner, Friedemann and Mahler, Günter},
	year = {2005},
	pages = {066118},
}

@article{kosloff_quantum_2014,
	title = {Quantum {Heat} {Engines} and {Refrigerators}: {Continuous} {Devices}},
	volume = {65},
	journal = {Annual Review of Physical Chemistry},
	author = {Kosloff, Ronnie and Levy, Amikam},
	year = {2014},
	pages = {365--393},
}

@article{roulet_autonomous_2018,
	title = {An autonomous single-piston engine with a quantum rotor},
	volume = {3},
	journal = {Quantum Science and Technology},
	author = {Roulet, Alexandre and Nimmrichter, Stefan and Taylor, Jacob M},
	year = {2018},
	pages = {035008},
}

@article{niedenzu_concepts_2019,
  title={Concepts of work in autonomous quantum heat engines},
  author={Niedenzu, Wolfgang and Huber, Marcus and Boukobza, Erez},
  journal={Quantum},
  volume={3},
  pages={195},
  year={2019},
}

@article{verteletsky_revealing_2020,
	title = {Revealing the strokes of autonomous quantum heat engines with work and heat fluctuations},
	volume = {101},
	journal = {Physical Review A},
	author = {Verteletsky, Katérina and Mølmer, Klaus},
	year = {2020},
	pages = {010101},
}

@article{rasola_proposal_2025,
	title={Proposal for an autonomous quantum heat engine},
	author={Miika Rasola and Vasilii Vadimov and Tuomas Uusnäkki and Mikko Möttönen},
	journal={SciPost Phys.},
	volume={19},
	pages={101},
	year={2025},
	publisher={SciPost},
	doi={10.21468/SciPostPhys.19.4.101},
}

@article{blais_cavity_2004,
	title = {Cavity quantum electrodynamics for superconducting electrical circuits: {An} architecture for quantum computation},
	volume = {69},
	journal = {Physical Review A—Atomic, Molecular, and Optical Physics},
	author = {Blais, Alexandre and Huang, Ren-Shou and Wallraff, Andreas and Girvin, Steven M and Schoelkopf, R Jun},
	year = {2004},
	pages = {062320},
}

@article{goppl_coplanar_2008,
	title = {Coplanar waveguide resonators for circuit quantum electrodynamics},
	volume = {104},
	journal = {Journal of Applied Physics},
	author = {Göppl, Martin and Fragner, A and Baur, M and Bianchetti, Romeo and Filipp, Stefan and Fink, Johannes M and Leek, Peter J and Puebla, G and Steffen, Lars and Wallraff, Andreas},
	year = {2008},
    pages = {113904},
}

@article{johansson_optomechanical-like_2014,
	title = {Optomechanical-like coupling between superconducting resonators},
	volume = {90},
	journal = {Physical Review A},
	author = {Johansson, J. R. and Johansson, G. and Nori, Franco},
	year = {2014},
	pages = {053833},
}

@article{rasola_low-characteristic-impedance_2024,
	title = {Low-characteristic-impedance superconducting tadpole resonators in the sub-gigahertz regime},
	volume = {6},
	journal = {Physical Review Research},
	author = {Rasola, Miika and Klaver, Samuel and Ma, Jian and Singh, Priyank and Uusnäkki, Tuomas and Suominen, Heikki and Möttönen, Mikko},
	year = {2024},
	pages = {043297},
}

@Article{Schmid1982,
  author    = {Albert Schmid},
  journal   = {Journal of Low Temperature Physics},
  title     = {On a quasiclassical Langevin equation},
  year      = {1982},
  pages     = {609--626},
  volume    = {49},
}

@book{pozar_microwave_2012,
	edition = {4th ed},
	title = {Microwave engineering},
	publisher = {Wiley},
	author = {Pozar, David M.},
	year = {2012},
}

@Article{Nakatsukasa2018,
  author    = {Yuji Nakatsukasa and Olivier S{\`{e}}te and Lloyd N. Trefethen},
  journal   = {{SIAM} Journal on Scientific Computing},
  title     = {The {AAA} Algorithm for Rational Approximation},
  year      = {2018},
  month     = {jan},
  number    = {3},
  pages     = {A1494--A1522},
  volume    = {40},
  doi       = {10.1137/16m1106122},
  publisher = {Society for Industrial {\&} Applied Mathematics ({SIAM})},
}

@Article{Vadimov2025,
  author    = {Vadimov, V. and Xu, M. and Stockburger, J. T. and Ankerhold, J. and Möttönen, M.},
  journal   = {Physical Review Research},
  title     = {Nonlinear-response theory for lossy superconducting quantum circuits},
  year      = {2025},
  issn      = {2643-1564},
  month     = mar,
  number    = {1},
  pages     = {013317},
  volume    = {7},
  doi       = {10.1103/physrevresearch.7.013317},
  publisher = {American Physical Society (APS)},
}

@article{perarnau-llobet_extractable_2015,
	title = {Extractable {Work} from {Correlations}},
	volume = {5},
	journal = {Physical Review X},
	author = {Perarnau-Llobet, Martí and Hovhannisyan, Karen V. and Huber, Marcus and Skrzypczyk, Paul and Brunner, Nicolas and Acín, Antonio},
	year = {2015},
	pages = {041011},
}

\section*{Acknowledgments}
% \vspace*{12pt}\noindent{\large\textbf{Acknowledgments}}\\ \noindent
We thank Heikki Suominen, Heidi Kivijärvi, Ashish Panigrahi, Qi-Ming Chen, Timm Mörstedt, Bayan Karimi, Jukka Pekola, Ilari Mäkinen, Christoforus Satrya, Mikko Tuokkola, Yoshiki Sunada, and Suman Kundu for discussions and help. 

\subsection*{Funding}
M.M. received funding from the Research Council of Finland Centre of Excellence program (project Nos. 352925 and 336810) and grant Nos. 316619 and 349594 (THEPOW), from the European Research Council under Advanced Grant No. 101053801 (ConceptQ), and from the Jane and Aatos Erkko Foundation through the project SystemQ (no project number). T.U. received personal funding from the Finnish Cultural Foundation (no project number). We also acknowledge the provision of facilities and technical support by Aalto University at OtaNano--Micronova Nanofabrication Centre.

\subsection*{Author contributions}
T.U. arranged the experimental setup, prepared the sample, conducted the experiments, carried out the data analysis, and wrote the manuscript with input from all the authors. M.R. designed the sample, prepared the theoretical background, and gave feedback and suggestions on the measurements. V.V provided theory support and carried out theoretical heat engine simulations. P.S. fabricated the sample chip and gave comments on the experiments. A.D. aided in carrying out the experiments. M.M. proposed the device concept and supervised the work in all respects.

\subsection*{Competing interests}
M.M. declares that he is a Co-Founder and Shareholder of quantum companies IQM Finland Oy and QMill Oy. The other authors declare no competing interests. 

\subsection*{Data and code availability}
The data and special codes that support the findings of this study are available from the corresponding author upon a reasonable request.
%are available at TBD. \href{https://doi.org/10.5281/zenodo.14935889}{https://doi.org/10.5281/zenodo.14935889}.

\clearpage
\section*{Methods}
% Main extra stuff for: 
% - Sample fabrication (Priyank)
% - More detailed experimental setup
% - Measurement protocols and IQ mixing etc.s
% - Thermal noise realizations: Ext. Fig. image of the heater
% - Heat engine theory/analysis (Miika and Vasilii): what to include, quantum vs quasiclassical description (old papers)

\subsection*{Sample fabrication}
The heat engine sample shown in Fig.~\ref{fig:setup}d is fabricated on a silicon wafer and cleaved into a $10\times10$\;mm chip using multiple micro- and nanofabrication steps. The used substrate is a high-purity, high-resistivity ($\rho > 11\ \text{k}\Omega\,\text{cm}$) silicon substrate with a thermally grown 300-nm layer of silicon oxide. The total substrate thickness is 675\;\textmu m. The ground plane, the CPW strips for all four resonators, and their respective drive lines are fabricated on the six-inch intrinsic-silicon wafer using optical lithography and reactive ion etching of a 200-nm film of sputtered niobium on top of the oxide layer. The widths of the CPW center conductors are $w = 10$\;\textmu m, and the gaps between the center conductor and the ground plane are $s = 6$\;\textmu m for all four resonators. 

For the driving resonator, we also fabricate a parallel-plate capacitor (PPC) in a tadpole resonator configuration~\cite{rasola_low-characteristic-impedance_2024}. For the parallel-plate capacitor, we first use atomic-layer deposition (ALD) to grow a 44-nm dielectric layer of $\mathrm{AlO}_x$ on the chip with 455 cycles in a $\mathrm{H_2O}$/TMA (trimethylaluminum) process at $200^{\circ}\text{C}$ in a Beneq TFS-500 system. Next, we protect the dielectric layer at the desired capacitor regions with AZ5214E resist and wet-etch the rest of the aluminium oxide away with a mixture of ammonium fluoride and hydrofluoric acid. Before depositing the top metal of the PPC, we first use argon milling to remove the intrinsic oxide from the niobium contact pad, ensuring proper galvanic contact. As the last step before lift-off, dicing, and bonding, we deposit a 50-nm aluminium layer in an electron beam evaporator for the PPC top metal.

The Josephson junctions in the coupling SQUID between the working-body and driving resonator are added using electron beam lithography (EBL) by adopting a refined recipe from the bridge-free fabrication of a large area Josephson junction mentioned in Ref.~\cite{zhang_bridge-free_2017}. The junction is made by depositing 30- and 45-nm aluminum layers using a two-angle shadow evaporation technique with in-situ oxidation for the insulating barrier in the SIS junction.
%spin-coating double-layer resist of a methyl methacrylate copolymer and polymethyl methacrylate, and then

\subsection*{Transmission measurements}
\begin{figure*}[t]
  \centering 
  \ifincludefigures\includegraphics[width=\linewidth]{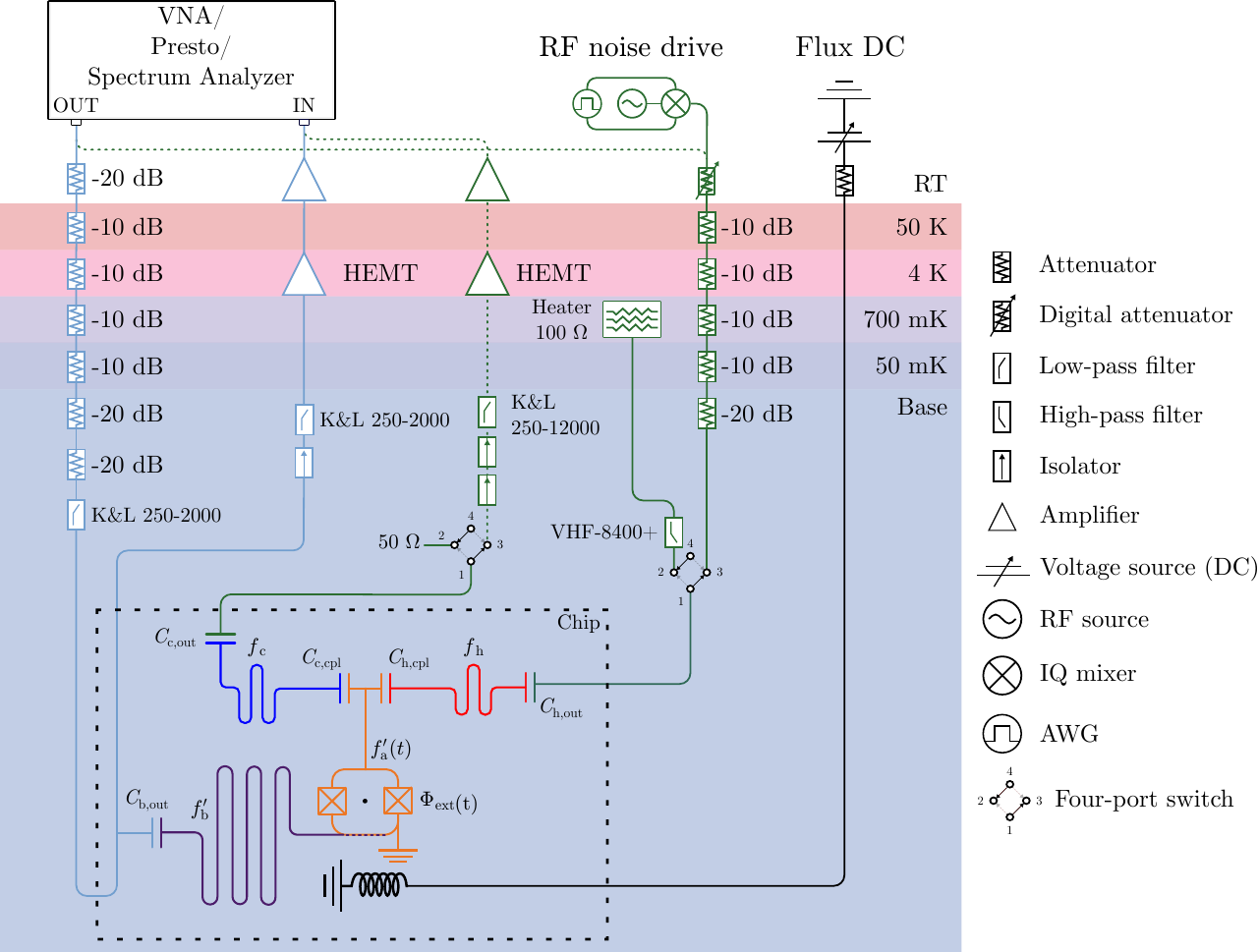}\fi
  \caption{\label{fig:ext_measurement_wiring}\textbf{Circuit and wiring diagram of the measurement setup.} Detailed circuit diagram of the measurement setup displaying the microwave drive lines and feedline with their corresponding attenuators and filters. Light blue circuitry depicts the low-frequency measurement setup for probing the driving resonator, and green circuitry depicts the high-frequency measurement setup for characterizing the filtering and working-body resonators and injecting the thermal noise into the hot filter.}
\end{figure*}

% Experimental setup, sample preparation etc.
The measurements in this work are carried out in a commercial dilution refrigerator cooled down to a base temperature of approximately 40\;mK. The heat engine chip is mounted on a copper sample holder with the on-chip microwave lines wire-bonded to the holder’s transmission lines using aluminum bonding wires. To suppress unwanted resonances, airbridges are formed by bonding wires between ground-plane sections across the waveguides on the chip. The sample holder is enclosed in a magnetic shield to protect the SQUID loop of the device from stray magnetic fields. The transmission line ports in the sample are connected to room-temperature microwave equipment via coaxial cables, attenuators, and amplifiers, as illustrated in the measurement wiring diagram in Fig.~\ref{fig:ext_measurement_wiring}. For low-frequency measurements, the signal lines to the sample include a total attenuation of 80 dB distributed across the various temperature stages of the cryostat, while the high-frequency RF noise line has a total attenuation of 60 dB. We use four-port switches both to select between different noise sources for the hot reservoir and to switch the high-frequency output line between a base-temperature termination and a room-temperature output.

% VNA measurements through feedline
We characterize and probe the resonators in the device using VNA transmission measurements, where we measure the S21 transmission coefficient with a Rohde~\&~Schwarz ZNB40 vector network analyzer. The reported probe power values used are the combination of the instrument output, the attenuation in the input lines, and the insertion losses of the various components, and we estimate a $\pm3$\;dBm uncertainty in the reported powers. For the high-frequency resonators, we probe through the input port of the hot filtering resonator and measure from the output port of the cold filtering resonators, thus probing the characteristic resonance frequencies of all of the three resonators in between. For the driving resonator, we probe the resonator using a feedline coupled to the inductive part of the tadpole resonator. For measuring the bare signal spectrum from the feedline, we use a Rohde~\&~Schwarz FSV40 spectrum analyzer.

% Time-domain measurements with Presto, IQ measurements
Time-domain trajectory averaging and single-shot measurements are carried out using a 16‑port microwave arbitrary waveform generator and analyzer (Presto, Intermodulation Products). Single‑trajectory measurements are implemented by sending 5\;\textmu s square probe pulses through the feedline to the driving resonator and measuring the transmitted signal. These pulses are generated by mixing an RF source with waveforms from an arbitrary waveform generator (AWG) in an IQ mixer. The transmitted signal is routed to the Presto readout card after the signal quadratures are separated with a second IQ mixer. The analog output is digitized, and the in‑phase (I) and quadrature (Q) components are integrated over a 5-\textmu s integration window for each shot. We acquire single shots by repeating the measurements thousands of times, from which we can get the trajectory‑averaged amplitude and phase by averaging the resulting single‑shot I and Q values.

\subsection*{Noise realizations}
To couple the working-body resonator to both cold and hot reservoirs, we implement photonic heat reservoirs for each filtering resonator. For the cold filter, the resonator is coupled via a drive line to the base temperature of the cryostat and terminated with a 50\;$\Omega$ load at the sample, as shown in Fig.~\ref{fig:ext_measurement_wiring}, effectively coupling it to the thermal noise floor of the cryostat. To establish the temperature gradient required for heat engine operation, the hot filter resonator is instead coupled via a transmission line to a noise source with an effective noise temperature higher than that of the cold filter. This hot reservoir is implemented in two ways: using an artificial RF noise source and a resistive heater.

% RF noise drive, mix noisy AWG signal with RF drive, repeat for millions of times
The quasithermal RF noise drive is implemented by generating a band-limited white-noise signal with an AWG and upconverting it using an IQ mixer driven by an RF source. We center the noise spectrum at the frequency of the hot filter resonator and choose a bandwidth three times the effective linewidth of the resonator, thereby approximating a white noise spectrum over the relevant frequency range, providing an effect essentially identical to that of a thermal source in a given temperature. We control the effective noise temperature of the RF noise by changing the power of the RF source signal. As the filter frequency is high and $k_\mathrm{B}T\sim h f_\mathrm{h}$, we model the noise temperature by inverting Planck's blackbody radiation law, giving the equation~\cite{pozar_microwave_2012}
\begin{equation}
    T_\mathrm{noise} = \frac{hf}{k_\mathrm{B}\ln\left(hfB/P_\mathrm{noise} + 1\right)},
\end{equation}
where $h$ is the Planck constant, $f$ is the center frequency of the noise spectrum, $k_\mathrm{B}$ is the Boltzmann constant, $B$ is the bandwidth of the noise, and $P_\mathrm{noise}$ is the power of the noise signal. 

\begin{figure}[t]
    \centering
    \includegraphics[width=0.6\linewidth]{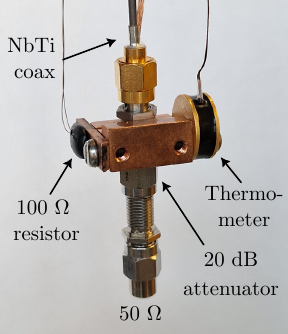}
    \caption{\textbf{Resistive blackbody heater used in the experiment}. The device consists of a 100-$\Omega$ resistor embedded in epoxy, a copper block as the thermal energy conduit, a thermometer, and the coaxial components for thermal radiation routing consisting of a 20-dB attenuator, 50-$\Omega$ terminator, and a coaxial NbTi cable.}
    \label{fig:ext_heater}
\end{figure}

% Heater -> 100 Ohm resistor with superconducting output line and thermometer for PID control
% Insertion loss: 10.4 K is high compared to Fig. 2f, but the line has a lot of insertion loss and we are not sure how much of the radiation is going through the NbTi line. Eccosorb 0.32 dB/GHz, DPDT switch 0.4 dBm, DC block 0.5 dB, VHF-8400+ max 2.5 dB
The resistive blackbody heater, displayed in Fig.~\ref{fig:ext_heater}, consists of superconducting DC wiring connected to a 100-$\Omega$ resistor embedded in epoxy, which is bonded to a copper block. The copper block clamps a 20-dB attenuator, the output of which is routed to the sample via a superconducting NbTi coaxial cable. A four-probe thermometer mounted on the copper block is used to monitor and calibrate its temperature. In operation, current through the 100-$\Omega$ resistor heats the copper block, which generates thermal radiation that is delivered through the superconducting coaxial cable to the hot-filter port of the sample via a four-port switch (see Fig.~\ref{fig:ext_measurement_wiring}). To use this device as a controlled thermal photon source, the thermometer signal is fed back to a PID controller that adjusts the current through the resistor to stabilize the copper block at a desired temperature. In this way, we can apply thermal photons with a tunable effective temperature to the sample. However, it is challenging to quantitatively compare the noise power and effective temperature between this heater and the RF noise source. The heater signal experiences substantial insertion loss due to the coupling between the resistor, copper block, and the coaxial line, as well as the attenuation from the IR filter, DC block, high-pass filter, and switch in the heater line. These losses account for the large discrepancy between the heater temperatures and the noise temperatures inferred from the measured RF noise power.

\subsection*{Heat engine analysis}

\subsubsection{Quasiclassical equation of motion}

For a quantitative model of the heat engine dynamics, we use the quasiclassical Langevin approach~\cite{Schmid1982, rasola_proposal_2025}. For the circuit shown in Fig.~\ref{fig:setup}b, we employ Kirchhoff's law taking account of thermal noise current. For each node of the circuit $j$ we assign flux $\varphi_j$. Consequently, the algebraic sum of currents in each of the nodes should vanish as follows:
\begin{equation}
    \sum_j \left(I_{j k} + \tilde I_{j k}\right) = 0,
\end{equation}
where $I_{j k}$ and $\tilde I_{j k}$ are noiseless and noisy parts of the current from the node $j$ to the node $k$, respectively. Depending on the element which connects nodes $j$ and $k$, the current is given by
\begin{align}
    I_{j k} &= C_{jk} \left (\ddot \varphi_j - \ddot \varphi_k \right)~\text{(capacitor)}, \\
    I_{j k} &= \frac{\varphi_j - \varphi_k - \Phi_{j k}}{L_{jk}}~\text{(inductor)}, \\
    I_{j k} &= I^{\mathrm{c}}_{jk} \sin \left[ \frac{2\pi}{\Phi_0} \left(\varphi_j - \varphi_k\right) \right]~\text{(Josephson junction)}, \\
    I_{j k} &= \frac{\dot \varphi_j - \dot \varphi_k}{R_{jk}}~\text{(resistor)}.
\end{align}
Here, $\Phi_{jk}$ is the external flux bias, $C_{jk}$, $L_{jk}$, $I^\mathrm{c}_{jk}$, and $R_{jk}$ are capacitance, inductance, critical current, and resistance of the respective elements which connect nodes $j$ and $k$. A resistor connecting nodes $j$ and $k$ also gives rise to the noise current source $\tilde I_{j k}$ with the power spectral density
\begin{equation}
    \int\limits_{-\infty}^{+\infty} \left \langle \tilde I_{jk}(t) \tilde I_{jk}(0) \right \rangle
    \mathrm{e}^{\mathrm{i} \omega t}\;\mathrm dt = \frac{\hbar \omega}{R_{jk}} \coth \left(\frac{\hbar \omega}{2 k_\mathrm B T_{jk}}\right),
    \label{eq:noise-psd}
\end{equation}
where $T_{jk}$ is the temperature of the respective resistor.

For lumped elements, listed above, current through the element is antisymmetric $I_{j k} = -I_{k j}$. However, we also have distributed elements in the circuit, namely CPW segments. Current through the CPW segment at the nodes $j$ and $k$ is time-nonlocal and cannot be expressed only through fluxes $\varphi_j$ and $\varphi_k$ and their derivatives. However, it can be expressed through derivatives of fluxes and auxiliary degrees of freedom $\chi_n$, $n = 1, 2, \ldots$ as
\begin{align}
    I_{j k} &= \frac{\pi \left(2 \ddot \varphi_{j} + \ddot \varphi_k\right)}{
        6 Z_{j k} \omega_{j k}
    } + \frac{
        \omega_{j k} \left(\varphi_j - \varphi_k\right)
    }{
        \pi Z_{j k}
    } + \sum\limits_{n=1}^{\infty} \frac{\ddot \chi_n}{n}, \\
    I_{k j} &= \frac{\pi \left(2 \ddot \varphi_{k} + \ddot \varphi_j\right)}{
        6 Z_{j k} \omega_{j k}
    } + \frac{
        \omega_{j k} \left(\varphi_k - \varphi_j\right)
    }{
        \pi Z_{j k}
    } + \sum\limits_{n=1}^{\infty} \frac{(-1)^n\ddot \chi_n}{n}.
\end{align}
Here, $Z_{j k}$ is the characteristic impedance of the CPW segment and $\omega_{j k}$ is its half-wavelength frequency. The auxiliary degrees of freedom $\chi_n$ in turn obey equations
\begin{equation}
    \ddot \chi_n + n^2 \omega_{j k}^2 \chi_n + 2 \frac{\ddot \varphi_j + (-1)^n \ddot \varphi_k}{
        \pi \omega_{j k} Z_{j k} n
    } = 0.
\end{equation}
Taking $N$ degrees of freedom into account provides quite an accurate approximation to the admittance of the CPW segment up to frequencies $\sim N \omega_{jk}$. In actual simulations, we keep only two auxiliary degrees of freedom for each of the CPW pieces in the circuit.

\begin{table}[h]
    \caption{\textbf{Parameters used in numerical calculations.} In relation to the model in Fig.~\ref{fig:setup}b, $\omega_x$ refers to the angular resonance frequency of the transmission line part of resonator $x\in\{\rm a,b,c,h\}$, $Z_x$ is the characteristic impedance of the transmission line part of $x$, $C_{x\rm, cpl}$ and $C_{x\rm, out}$ are the coupling and output capacitances of the resonators, $I_1$ and $I_2$ are the critical currents of the SQUID junctions, $L_\mathrm{cpl}$ is the galvanic inductive coupling to the SQUID, $C_\mathrm{PPC}$ is the parallel-plate capacitance, $T_\mathrm{h}$ and $T_\mathrm{c}$ are the temperatures of the hot and cold resistors, respectively, $R_y$ is the resistance of the coupled resistor for $y\in\{\rm b,c,h\}$, $Z_0$ is the characteristic impedance of the feedline, and $T_\mathrm{base}$ is the base temperature of the cryostat.}
    
    \begin{tabular}{cccc}
    \hline %\hline
        Parameter & Value & Parameter & Value\\
        \hline
        $\omega_\mathrm h / (2\pi)$ & $10.69$~GHz &
        $I_1$ & $1.39~\mu\textrm{A}$ \\
        % \hline
        $Z_\mathrm h$ & $50~\Omega$ &
        $I_2$ & $1.44~\mu\textrm{A}$ \\
        % \hline
        $C_{\mathrm h, \mathrm{out}}$ & $25.86$~fF &
        $L_\mathrm{cpl}$ & $0.091$~nH \\
        % \hline
        $\omega_\mathrm c / (2\pi)$ & $9.37$~GHz &
        $\omega_\mathrm b / (2\pi )$ & $15.0$~GHz \\
        % \hline
        $Z_\mathrm c$ & $50~\Omega$ &
        $Z_\mathrm b$ & $50~\Omega$ \\
        % \hline
        $C_{\mathrm c, \mathrm{out}}$ & $30.0$~fF &
        $R_\mathrm b$ & $42.24~\mathrm{k}\Omega$ \\
        % \hline
        $T_\mathrm c$ & $20$~mK &
        $C_\mathrm{PPC}$ & $144.2$~pF \\
        % \hline
        $C_{\mathrm h, \mathrm{cpl}}$ & $11.48$~fF &
        $C_{\mathrm b,\mathrm{out}}$ & $29.09$~fF \\
        % \hline
        $C_{\mathrm c, \mathrm{cpl}}$ & $10.52$~fF &
        $T_\mathrm b$ & $20$~mK \\
        % \hline
        $\omega_{\mathrm a} / (2 \pi)$ & $6.12$~GHz &
        $Z_0$ & $50~\Omega$ \\
        % \hline
        $Z_\mathrm a$ & $50~\Omega$ &
        $T_\mathrm{base}$ & $20$~mK \\
        \hline
        % \hline
    \end{tabular}
    \label{tab:num_parameters}
\end{table}

The last step is unraveling of the colored noise $\tilde I_{j k}$. Again, for each noise source we introduce auxiliary degrees of freedom $\eta_n$, $n = 1, 2, \ldots$, which obey equations
\begin{equation}
    \dot \eta_n = -\nu_n \eta_n + \xi_n(t),
\end{equation}
where $\nu_n$ are certain decay rates to be determined and $\xi_n(t)$ are independent delta-correlated noise sources $\langle \xi_{n}(t) \xi_{n'}(t') \rangle = \delta_{n n'} \delta(t - t')$. We adopt the following approximation for the current noise
\begin{equation}
    \tilde I_{j k} = -\tilde I_{k j} = c_0 \xi_0(t) + \sum_{n=1}^{\infty} c_n \dot \eta_n(t).
    \label{eq:matsubara}
\end{equation}
Here $\xi_0(t)$ is yet another white noise source. The parameters $c_0$, $c_n$, and $\nu_n$ are adjusted in such a way that Eq.~\eqref{eq:noise-psd} is fulfilled:
\begin{equation}
    c_0^2 + \sum\limits_{n=1}^{\infty} \frac{c_n^2 \omega^2}{\omega^2 + \nu_n^2} = \frac{\hbar \omega}{R_{j k}} \coth \left(\frac{\hbar \omega}{2 k_\mathrm B T_{jk}}\right).
\end{equation}
This is valid for $c_0^2 = c_n^2 / 2 = 2 k_\mathrm B T_{jk} / R_{j k}$ and $\nu_n =~2\pi n k_\mathrm B T_{j k} / \hbar$. However, direct truncation of the series~\eqref{eq:matsubara} provides poor convergence, and in practice it is more efficient to construct a rational approximation of $\coth(z)$ numerically using the adaptive Antoulas--Anderson algorithm~\cite{Nakatsukasa2018, Vadimov2025}. This way, we can approximate current noise using much fewer auxiliary degrees of freedom.

\subsubsection{Gaussian approximation}
If we collect all the equations for all the degrees of freedom, physical and auxiliary, we obtain a system which, in general, has the following form:
\begin{equation}
\frac{\mathrm d \boldsymbol{x}}{\mathrm d t} = 
\boldsymbol{A} \boldsymbol{x} - \boldsymbol x_\mathrm{ext} + \sum_j \boldsymbol{p}_j
\sin \left(\boldsymbol{q}_j^\mathsf T \boldsymbol{x}\right) + \boldsymbol B \boldsymbol \xi(t).
\end{equation}
Here, vector $\boldsymbol x$ contains all the degrees of freedom: $\varphi_j$, $\dot \varphi_j$, $\chi_n$, $\dot \chi_n$, and $\eta_n$. Constant matrix $\boldsymbol A$ describes linear couplings between these degrees of freedom, constant vector $\boldsymbol x_\mathrm{ext}$ is introduced to describe the flux bias, vectors $\boldsymbol p_j$ and $\boldsymbol q_j$ describe the current through the Josephson junctions, vector $\boldsymbol \xi(t)$ is a vector of independent delta-correlated Gaussian noise sources, and matrix $\boldsymbol B$ provides coupling between the noise terms and dynamical degrees of freedom. To analyze this equation, we introduce a Gaussian approximation. We expand $\boldsymbol x(t) = \langle \boldsymbol x(t) \rangle - \tilde{\boldsymbol x}(t)$ and assume $\tilde{\boldsymbol x}(t)$ to be zero-mean Gaussian process. Here, the averaging is with respect to noise realizations. Under this assumption, we can write a closed system of equations on $\langle \boldsymbol x(t) \rangle$ and covariance matrix $\boldsymbol C(t) := \langle \tilde{\boldsymbol x}(t) \tilde{\boldsymbol x}^\mathsf T(t) \rangle$:
\begin{equation}
    \frac{\mathrm d \langle \boldsymbol x \rangle}{\mathrm d t} = 
    \boldsymbol A \langle \boldsymbol x \rangle - \boldsymbol x_\mathrm{ext} + \sum\limits_j
    \boldsymbol p_j \sin \left(
    \boldsymbol q_j^\mathsf T \langle \boldsymbol x \rangle
    \right) \mathrm{e}^{
    -\frac{\boldsymbol q_j^\mathsf T
    \boldsymbol C \boldsymbol q_j}{2} 
    },
    \label{eq:mean}
\end{equation}
\begin{multline}
    \frac{\mathrm d \boldsymbol C}{\mathrm d t} = 
    \boldsymbol A \boldsymbol C + \boldsymbol C \boldsymbol A^\mathsf T + \boldsymbol B
    \boldsymbol B^\mathsf T
    \\+ \sum\limits_j
    \left(
    \boldsymbol p_j \boldsymbol q_j^\mathsf T \boldsymbol C + 
    \boldsymbol C \boldsymbol q_j \boldsymbol p_j^\mathsf T
    \right)\cos \left(
    \boldsymbol q_j^\mathsf T \langle \boldsymbol x \rangle
    \right) \mathrm{e}^{
    -\frac{\boldsymbol q_j^\mathsf T
    \boldsymbol C \boldsymbol q_j}{2} 
    }.
    \label{eq:covariance}
\end{multline}
These equations present a system of ordinary differential equations for mean values and the covariance matrix, which are feasible for numerical methods.

\subsubsection{Numerical results}

\begin{figure*}[t]
    \includegraphics[width=\linewidth]{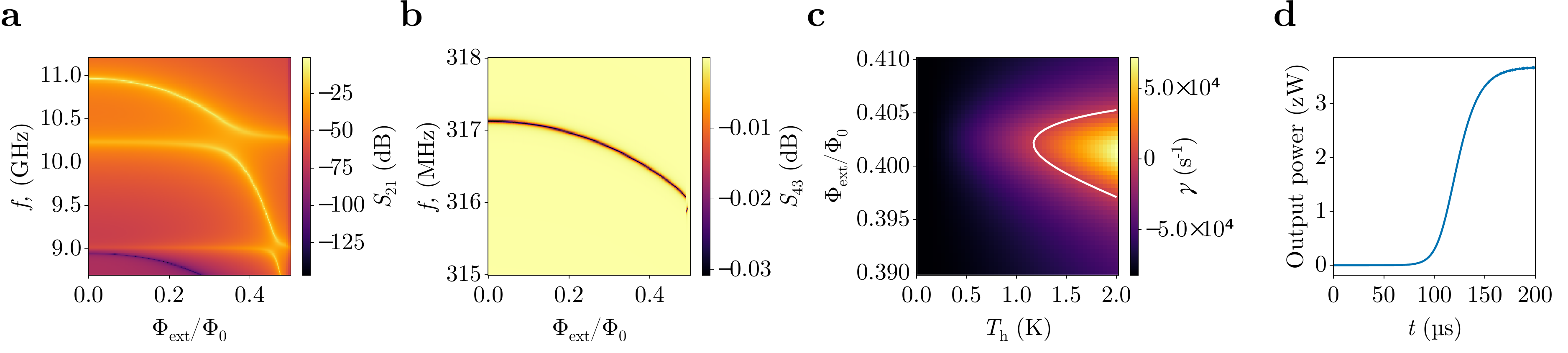}
    \caption{
    \label{fig:numerics}
    \textbf{Quasiclassical simulations.} \textbf{a}, \textbf{b}, Magnitude of the transmission coefficient from (\textbf{a}) hot to cold filter ports and (\textbf{b}) ports of the input feedline as functions of the normalized external flux and probe frequency. \textbf{c}, Maximum growth rate of the linearized modes of the driving resonator as a function of the temperature of the hot reservoir and normalized external flux. The area with a positive rate is surrounded by a white curve and corresponds to the engine operation starting from arbitrary small fluctuations. \textbf{d}, Output power of the heat engine as a function of time. The solution shows the start of the heat engine operation from a state close to the stationary state. The temperature of the hot reservoir is $T_\mathrm h = 2$~K and flux bias is $\Phi_\mathrm{ext}/\Phi_0 = 0.4015$. Circuit parameters used for the calculations are given in Table~\ref{tab:num_parameters}.
    }
\end{figure*}

% \begin{figure}[t]
%   \centering 
%   \ifincludefigures\includegraphics[width=\linewidth]{increment.pdf}\fi
%   \caption{\label{fig:increment}}
% \end{figure}

First, we study the dynamics of the engine in the vicinity of the stationary state. For a given external flux $\Phi_\mathrm{ext}$ and a temperature of the hot reservoir $T_\mathrm h$, we find a stationary solution $\boldsymbol x(t) = \boldsymbol x_0(\Phi_\mathrm{ext}, T_\mathrm h)$ and $\boldsymbol C(t) = \boldsymbol C_0(\Phi_\mathrm{ext}, T_\mathrm h)$, and linearize the equations of motion around this point. Using the linearized equations, we can numerically reproduce spectroscopic measurements of the engine shown in Figs.~\ref{fig:characterization}a and \ref{fig:characterization}b. The results of the numerical analysis are presented in Figs.~\ref{fig:numerics}a and \ref{fig:numerics}b. We use the numerically evaluated transmission coefficients to adjust the parameters of the model circuit, see Table~\ref{tab:num_parameters} for the results. 

After we estimate the parameters of the model, we proceed with stability analysis of the stationary solution. The theory predicts that with the increase of the hot-reservoir temperature, the decay rate of the fundamental mode of the driving resonator decreases and eventually changes its sign~\cite{rasola_proposal_2025}. A change of the sign indicates a dynamical instability of the stationary state of the engine, which leads to the start of its operation from arbitrarily small fluctuations. In Fig.~\ref{fig:numerics}c, we show the instability increment of the fundamental mode of the driving resonator. We observe that our numerical analysis confirms theoretical predictions: for a certain flux bias, the engine starts autonomous operation from the stationary state if the temperature of the hot reservoir is sufficiently high. 

Finally, we simulate the real-time dynamics of the heat engine. We analyze the case where the engine operation starts from the small fluctuations. We pick the temperature of the hot reservoir and the flux bias from the unstable parameter area shown in Fig.~\ref{fig:numerics}c and solve Eqs.~\eqref{eq:mean} and~\eqref{eq:covariance} for the initial conditions close to the stationary state. In Fig.~\ref{fig:numerics}d, we show the output power of the heat engine as a function of time. We clearly observe that the engine starts to generate power from almost a stationary state, and eventually the engine reaches the operational mode. We associate the large discrepancy in the simulated and measured output powers with a parameter mismatch, probably due to too weak coupling between the feedline and the driving resonator in the simulations.
 
\end{document}